\documentclass[twocolumn]{aastex62}
\pdfoutput=1 
\usepackage{amsmath,amstext}
\usepackage[T1]{fontenc}
\usepackage[figure,figure*]{hypcap}
\graphicspath{{./}{figures/}}

\received{?}
\revised{?}
\accepted{?}
\submitjournal{AAS}

\shorttitle{TOI-2018}
\shortauthors{Dai et al.}

\begin{document}

\title{A Mini-Neptune Orbiting the Metal-poor K Dwarf BD+29 2654}

\author[0000-0002-8958-0683]{Fei Dai}
\altaffiliation{fdai@caltech.edu}
\altaffiliation{NASA Sagan Fellow}
\affiliation{Division of Geological and Planetary Sciences,
1200 E California Blvd, Pasadena, CA, 91125, USA}
\affiliation{Department of Astronomy, California Institute of Technology, Pasadena, CA 91125, USA}

\author[0000-0001-5761-6779]{Kevin C.\ Schlaufman}
\affiliation{William H.\ Miller III Department of Physics and Astronomy,
Johns Hopkins University, 3400 N Charles St, Baltimore, MD 21218, USA}

\author[0000-0001-6533-6179]{Henrique Reggiani}
\affiliation{The Observatories of the Carnegie Institution for Science, 813 Santa Barbara St, Pasadena, CA 91101, USA}

\author[0000-0002-0514-5538]{Luke Bouma}
\altaffiliation{51 Pegasi b Fellow}
\affiliation{Department of Astronomy, California Institute of Technology, Pasadena, CA 91125, USA}

\author[0000-0001-8638-0320]{Andrew W. Howard}
\affiliation{Department of Astronomy, California Institute of Technology, Pasadena, CA 91125, USA}

\author[0000-0003-1125-2564]{Ashley Chontos}
\affiliation{Department of Astrophysical Sciences, Princeton University, 4 Ivy Lane, Princeton, NJ 08540, USA}
\affiliation{Henry Norris Russell Fellow}
\affiliation{Institute for Astronomy, University of Hawai`i, 2680 Woodlawn Drive, Honolulu, HI 96822, USA}

\author[0000-0001-9771-7953]{Daria Pidhorodetska} 
\affiliation{Department of Earth and Planetary Sciences, University of California, Riverside, CA 92521, USA}

\author[0000-0002-4290-6826]{Judah Van Zandt}
\affiliation{Department of Physics \& Astronomy, University of California Los Angeles, Los Angeles, CA 90095, USA}

\author[0000-0001-8898-8284]{Joseph M. Akana Murphy}
\altaffiliation{NSF Graduate Research Fellow}
\affiliation{Department of Astronomy and Astrophysics, University of California, Santa Cruz, CA 95064, USA}

\author[0000-0003-3856-3143]{Ryan A. Rubenzahl}
\altaffiliation{NSF Graduate Research Fellow}
\affiliation{Department of Astronomy, California Institute of Technology, Pasadena, CA 91125, USA}

\author[0000-0001-7047-8681]{Alex S. Polanski}
\affil{Department of Physics and Astronomy, University of Kansas, Lawrence, KS 66045, USA}

\author[0000-0001-8342-7736]{Jack Lubin}
\affiliation{Department of Physics \& Astronomy, University of California Irvine, Irvine, CA 92697, USA}

\author[0000-0001-7708-2364]{Corey Beard}
\affiliation{Department of Physics \& Astronomy, University of California Irvine, Irvine, CA 92697, USA}

\author[0000-0002-8965-3969]{Steven Giacalone}
\affil{Department of Astronomy, University of California Berkeley, Berkeley, CA 94720, USA}

\author[0000-0002-5034-9476]{Rae Holcomb}
\affiliation{Department of Physics \& Astronomy, University of California Irvine, Irvine, CA 92697, USA}

\author[0000-0002-7030-9519]{Natalie M. Batalha}
\affiliation{Department of Astronomy and Astrophysics, University of California, Santa Cruz, CA 95060, USA}

\author{Ian Crossfield}
\affiliation{Department of Physics and Astronomy, University of Kansas, Lawrence, KS, USA}

\author{Courtney Dressing}
\affiliation{501 Campbell Hall, University of California at Berkeley, Berkeley, CA 94720, USA}

\author[0000-0003-3504-5316]{Benjamin Fulton}
\affiliation{NASA Exoplanet Science Institute/Caltech-IPAC, MC 314-6, 1200 E California Blvd, Pasadena, CA 91125, USA}

\author[0000-0001-8832-4488]{Daniel Huber}
\affiliation{Institute for Astronomy, University of Hawai`i, 2680 Woodlawn Drive, Honolulu, HI 96822, USA}
\affiliation{Sydney Institute for Astronomy, School of Physics, University of Sydney NSW 2006, Australia}

\author[0000-0002-0531-1073]{Howard Isaacson}
\affiliation{501 Campbell Hall, University of California at Berkeley, Berkeley, CA 94720, USA}
\affiliation{Centre for Astrophysics, University of Southern Queensland, Toowoomba, QLD, Australia}

\author[0000-0002-7084-0529]{Stephen R. Kane}
\affiliation{Department of Earth and Planetary Sciences, University of California, Riverside, CA 92521, USA}

\author[0000-0003-0967-2893]{Erik A. Petigura}
\affiliation{Department of Physics \& Astronomy, University of California Los Angeles, Los Angeles, CA 90095, USA}

\author[0000-0003-0149-9678]{Paul Robertson}
\affiliation{Department of Physics \& Astronomy, University of California Irvine, Irvine, CA 92697, USA}

\author[0000-0002-3725-3058]{Lauren M. Weiss}
\affiliation{Department of Physics and Astronomy, University of Notre Dame, Notre Dame, IN 46556, USA}

\author[0000-0003-2228-7914]{Alexander A. Belinski}
\affiliation{Sternberg Astronomical Institute Lomonosov Moscow State University 119992, Moscow, Russia, Universitetskii prospekt}

\author[0000-0001-6037-2971]{Andrew W. Boyle}
\affiliation{Department of Astronomy, California Institute of Technology, 1200 E. California Blvd, Pasadena, CA 91125 USA}

\author[0000-0002-7754-9486]{Christopher~J.~Burke}
\affiliation{Department of Physics and Kavli Institute for Astrophysics and Space Research, Massachusetts Institute of Technology, Cambridge, MA 02139, USA}

\author[0000-0001-7439-3618]{Amadeo Castro-González}
\affiliation{Centro de Astrobiolog\'ia (CAB, CSIC-INTA), Depto. de Astrof\'isica, ESAC campus, 28692, Villanueva de la Ca\~nada (Madrid), Spain}

\author[0000-0002-5741-3047]{David~ R.~Ciardi}
\affiliation{Caltech/IPAC-NASA Exoplanet Science Institute, 770 S. Wilson Avenue, Pasadena, CA 91106, USA}

\author[0000-0002-6939-9211]{Tansu Daylan}
\affiliation{Department of Astrophysical Sciences, Princeton University, Princeton, NJ 08544, USA}
\affiliation{LSSTC Catalyst Fellow}

\author[0000-0002-4909-5763]{Akihiko Fukui}
\affiliation{Komaba Institute for Science, The University of Tokyo,
3-8-1 Komaba, Meguro, Tokyo 153-8902, Japan}
\affiliation{Instituto de Astrof\'{i}sica de Canarias (IAC), 38205 La
Laguna, Tenerife, Spain}

\author[0000-0001-6171-7951]{Holden Gill}
\affil{Department of Astronomy, University of California Berkeley, Berkeley, CA 94720, USA}

\author[0000-0002-5169-9427]{Natalia~M.~Guerrero}
\affiliation{Department of Astronomy, University of Florida, Gainesville, FL, 32611, USA}
\affiliation{Department of Physics and Kavli Institute for Astrophysics and Space Research, Massachusetts Institute of Technology, Cambridge, MA 02139, USA}

\author[0000-0002-3439-1439]{Coel Hellier}
\affiliation{Astrophysics Group, Keele University, Staffordshire ST5 5BG, U.K}

\author[0000-0002-2532-2853]{Steve~B.~Howell}
\affiliation{NASA Ames Research Center, Moffett Field, CA 94035, USA}

\author[0000-0003-3742-1987]{Jorge Lillo-Box}
\affiliation{Centro de Astrobiolog\'ia (CAB, CSIC-INTA), Depto. de Astrof\'isica, ESAC campus, 28692, Villanueva de la Ca\~nada (Madrid), Spain}

\author[0000-0001-9087-1245]{Felipe Murgas}
\affiliation{Instituto de Astrofísica de Canarias (IAC), Calle Vía Láctea s/n, 38205 La Laguna, Tenerife, Spain }
\affiliation{Departamento de Astrof\'isica, Universidad de La Laguna (ULL), E-38206 La Laguna, Tenerife, Spain}

\author[0000-0001-8511-2981]{Norio Narita}
\affiliation{Komaba Institute for Science, The University of Tokyo,
3-8-1 Komaba, Meguro, Tokyo 153-8902, Japan}
\affiliation{Astrobiology Center, 2-21-1 Osawa, Mitaka, Tokyo 181-8588, Japan}
\affiliation{Instituto de Astrof\'{i}sica de Canarias (IAC), 38205 La
Laguna, Tenerife, Spain}

\author[0000-0003-0987-1593]{Enric Pallé}
\affiliation{Instituto de Astrofísica de Canarias (IAC), Calle Vía Láctea s/n, 38205 La Laguna, Tenerife, Spain }

\author{David~R.~Rodriguez}
\affiliation{Space Telescope Science Institute, 3700 San Martin Drive, Baltimore, MD, 21218, USA}

\author[0000-0002-2454-768X]{Arjun B. Savel}
\affiliation{Department of Astronomy, University of Maryland, College Park, College Park, MD 20742 USA}

\author[0000-0002-1836-3120]{Avi~Shporer}
\affiliation{Department of Physics and Kavli Institute for Astrophysics and Space Research, Massachusetts Institute of Technology, Cambridge, MA 02139, USA}

\author[0000-0002-3481-9052]{Keivan G.\ Stassun}
\affiliation{Department of Physics and Astronomy, Vanderbilt University, Nashville, TN 37235, USA}

\author[0009-0008-5145-0446]{Stephanie Striegel}
\affiliation{SETI Institute, 339 N Bernardo Ave Suite 200, Mountain View, CA 94043, USA}
\affiliation{NASA Ames Research Center, Moffett Field, CA 94035, USA}

\author[0000-0002-4715-9460]{Douglas A. Caldwell}
\affiliation{NASA Ames Research Center, Moffett Field, CA 94035, USA}

\author[0000-0002-4715-9460]{Jon M. Jenkins}
\affiliation{NASA Ames Research Center, Moffett Field, CA 94035, USA}

\author[0000-0003-2058-6662]{George R. Ricker}
\affiliation{Department of Physics and Kavli Institute for Astrophysics and Space Research, Massachusetts Institute of Technology, Cambridge, MA 02139, USA}

\author[0000-0002-6892-6948]{Sara Seager}
\affiliation{Department of Physics and Kavli Institute for Astrophysics and Space Research, Massachusetts Institute of Technology, Cambridge, MA
02139, USA}
\affiliation{Department of Earth, Atmospheric and Planetary Sciences, Massachusetts Institute of Technology, Cambridge, MA 02139, USA}
\affiliation{Department of Aeronautics and Astronautics, MIT, 77 Massachusetts Avenue, Cambridge, MA 02139, USA}

\author[0000-0001-6763-6562]{Roland Vanderspek}
\affiliation{Department of Physics and Kavli Institute for Astrophysics and Space Research, Massachusetts Institute of Technology, Cambridge, MA 02139, USA}

\author[0000-0002-4265-047X]{Joshua N. Winn}
\affiliation{Department of Astrophysical Sciences, Princeton University, 4 Ivy Lane, Princeton, NJ 08544, USA}



\begin{abstract}
\noindent We report the discovery and Doppler mass measurement
of a 7.4-day 2.3-$R_\oplus$ mini-Neptune around a metal-poor K dwarf BD+29 2654
(TOI-2018). Based on a high-resolution Keck/HIRES spectrum, the Gaia parallax, and multi-wavelength photometry from the ultraviolet to
the mid-infrared, we found that the host star has $T_{\text{eff}}=4174^{+34}_{-42}$ K, $\log{g}=4.62^{+0.02}_{-0.03}$, $[\text{Fe/H}]=-0.58\pm0.18$, $M_{\ast}=0.57\pm0.02~M_{\odot}$, and $R_{\ast}=0.62\pm0.01~R_{\odot}$. Precise Doppler measurements with Keck/HIRES revealed a planetary mass of $M_{\text{p}}=9.2\pm2.1~M_{\oplus}$ for TOI-2018 b. TOI-2018 b has a mass and
radius that are consistent with an Earth-like core with a $\sim1\%$-by-mass hydrogen/helium envelope, or an ice-rock mixture. The mass of TOI-2018 b is close to the threshold for run-away accretion and hence giant planet formation. Such a threshold is predicted to be around 10$M_\oplus$ or lower for a low-metallicity (low-opacity) environment. If TOI-2018 b is a planetary core that failed to undergo run-away accretion, it may underline the reason why giant planets are rare around low-metallicity host stars (one possibility is their shorter disk lifetimes).   With a K-band magnitude of 7.1, TOI-2018 b may be a suitable target for transmission spectroscopy with the James Webb Space Telescope. The system is also amenable to metastable Helium observation; the detection of a Helium exosphere would help distinguish between a H/He enveloped planet and a water world. 

\end{abstract}


\section{Introduction}
The occurrence rate of giant planets is strongly correlated with the host star metallicity \citep{Gonzalez,Santos2004,Fischer2005}. This strong correlation has been regarded as supporting evidence for the core accretion theory of planet formation \citep{POLLACK199662}. In a metal-rich disk, a high abundance of solid materials helps planet embryos grow quickly to the critical mass that initiates run-away gas accretion and gas giant formation. On the other hand, the close-in, sub-Neptune planets (<1AU, 1-4$R_\oplus$, also known as the Kepler-like planets) are much more common around solar-type stars compared to giant planets \citep[e.g.][]{Howard+2012,Petigura2013,Fressin}. The occurrence rate of sub-Neptune planets also has a much weaker dependence on host metallicity compared to the giant planets \citep[e.g.][]{WangFischer,Petigura_CKS}. One might argue that the formation of gas giant planets may be a threshold-crossing event that has to occur when the gaseous disk is still present. On the other hand, the formation of sub-Neptune planets is less demanding on core assembly rate and can proceed in low-metallicity environments. The super-Earths planets may not accrete a substantial envelope \citep{Rogers}, whereas the mini-Neptunes only acquire their envelopes towards the final vestige of the disk when the disk starts to become optically thin \citep{Lee2016,2018MNRAS.476.2199L}.

So far, most planet occurrence studies \citep[e.g. {\it Kepler} ][]{Borucki+2011} are based on surveys of stars with sun-like metallicities [Fe/H] between $-0.5$ and $0.5$. It is not clear how the occurrence-metallicity correlation extends to a more metal-depleted regime ([Fe/H]\,$<-0.5$). The standard minimum mass solar nebula \citep{Hayashi} has about 30$M_\oplus$ of solid materials within the innermost 1~AU for the {\it in-situ} formation of the close-in sub-Neptune planets. The total amount of solid materials would decrease to 3-9 $M_\oplus$ for -1<[Fe/H]<-0.5. Such a limited supply of solids may prevent the formation of multiple close-in sub-Neptune planets.

To further investigate the influence of host metallicity, we (Schlaufman et al., in prep) are studying the occurrence rate of transiting sub-Neptune planets around metal-poor (-1<[Fe/H]<-0.5) stars observed as the {\it TESS} mission \citep{Ricker}. Thanks to the nearly full-sky coverage of {\it TESS}, we were able to cross-match the {\it TESS} Input Catalog \citep{Stassun} with ground-based spectroscopic surveys, including the Large Sky Area Multi-Object Fiber Spectroscopic Telescope (LAMOST) Low-Resolution Survey (LRS) Data Release (DR) 6 \citep{Lamost}, the Radial Velocity Experiment (RAVE) DR6 \citep{Steinmetz}, and the GALactic Archaeology with Hermes (GALAH) DR3 \citep{Buder}. We identified a sample of about 10{,}000 dwarf stars with -1<[Fe/H]<-0.5. We have carried out a systematic search for transiting planets among this sample. This search led to the discovery of BD+29 2654, which was also discovered independently by the {\it TESS} team  as TOI-2018 \citep{Guerrero}. TOI-2018 is a bright, nearby K-dwarf among our transiting planet hosts that is particularly amenable to follow-up observations. We also present an additional transiting signal, TOI-2018.02, which was not reported by the {\it TESS} team \citep{Guerrero} due to its lower signal-to-noise ratio (SNR). We carried out a detailed characterization of the host star and Doppler mass measurements of the planets to help the community plan follow-up observations of this system. 

The paper is organized as follows. Section 2 presents a detailed characterization of the host star TOI-2018 with particular attention to its metallicity. Section 3 describes our transit detection and modeling of this system based on the light curves. In Section 4, we present the radial velocity (RV) measurements of TOI-2018 and the resultant constraints on the planetary masses. Section 5 discusses the implications of our findings.

\section{Stellar Properties}\label{sec:stellar_para}

\subsection{Fundamental and Photospheric Parameters}
\label{sec:star}
To derive the stellar parameters, we analyzed an archival high-resolution, high-SNR spectra of TOI-2018 taken with the High Resolution Echelle Spectrometer on the 10-meter Keck Telescope \citep[Keck/HIRES,][]{HIRES} on Jul 26 2012. The spectrum was taken without the iodine-cell and reached an SNR of about 200 per reduced pixel near 5500\AA~after a 600-sec exposure. 
We made use of both spectroscopy and 
isochrones to infer the photospheric
and fundamental stellar parameters as described in \citet{reggiani2022}.  Isochrones are especially useful
for determining the effective temperature $T_{\text{eff}}$ of the star, because
high-quality multi-wavelength photometry from the ultraviolet to the red
optical is available.
Similarly, the availability of the Gaia DR3 parallax-based
distance of TOI-2018 makes the calculation of surface gravity $\log{g}$
via isochrones more straightforward than it has
traditionally been.  With good
constraints on both $T_{\text{eff}}$ and $\log{g}$
from isochrone fitting, the equivalent widths of iron lines can
be used to determine metallicity $[\text{Fe/H}]$
and microturbulence $\xi$ by minimizing the dependence of individual
line-based iron abundance inferences on their reduced equivalent width. 


\begin{deluxetable}{lcc}
\tablecaption{Adopted Stellar Parameters}
\tablewidth{0pt}
\tablehead{
\colhead{Property} & \colhead{Value} & \colhead{Unit}}
\startdata
\label{tab:stellar_param_table}
SDSS DR13 $u$ & $13.385\pm0.0264$ & AB mag \\
Gaia DR2 DR2 $G$   & $9.720\pm0.002$ & Vega mag \\
2MASS $J$ & $7.844\pm0.021$ & Vega mag \\ 
2MASS $H$ & $7.255\pm0.020$ & Vega mag \\ 
2MASS $K$ & $7.104\pm0.017$ & Vega mag \\ 
WISE W1 & $6.984\pm0.062$   & Vega mag \\ 
WISE W2 & $7.071\pm0.020$   & Vega mag \\ 
Gaia DR3 parallax & $35.666\pm0.014$ & mas \\
\hline
\multicolumn{3}{l}{\textbf{Spectroscopically inferred parameters}} \\
$[\text{Fe/H}]$ &  $-0.58\pm0.18$ & \\
$v$sin$i$ &  $<2$ & km s$^{-1}$ \\
$S_{HK}$ & $0.92 \pm 0.03$ &\\
log$R_{\rm HK}^\prime$ & -4.75 $\pm$ 0.04&\\
\hline
\multicolumn{3}{l}{\textbf{Isochrone-inferred parameters}} \\
Effective temperature $T_{\text{eff}}$ & $4174^{+34}_{-42}$ & K \\
Surface gravity $\log{g}$ & $4.62^{+0.02}_{-0.03}$ & cm s$^{-2}$ \\
Stellar mass $M_{\ast}$ & $0.57 \pm 0.02$ & $M_{\odot}$ \\
Stellar radius $R_{\ast}$ & $0.62\pm0.01$ & $R_{\odot}$ \\
Luminosity $L_{\ast}$ & $0.10 \pm 0.01$ & $L_{\odot}$ \\
Distance & $28.038\pm0.011$ & pc \\
\enddata
\tablecomments{Substantial systematic uncertainties may exist between different isochronal models \citep{Tayar}.} 
\end{deluxetable}

The inputs to our photospheric and fundamental stellar parameter
inference are the equivalent widths of \ion{Fe}{1} and \ion{Fe}{2}
atomic absorption lines, multiwavelength photometry, the Gaia parallax,
and an extinction estimate.  Using atomic absorption line data from
\cite{galarza2019} for lines that are relatively
insensitive to stellar activity \citep{melendez2014}, we measured the equivalent
widths by fitting Gaussian profiles with the \texttt{splot} task in
\texttt{IRAF} to our continuum-normalized spectrum. We also 
confirmed our \texttt{splot} equivalent widths by remeasuring the 
lines using \texttt{iSpec} \citep{blanco-cuaresma2014,blanco-cuaresma2019}. 
We only compared the clean (unblended) lines with our \texttt{splot} 
measurements and we concluded that there were no substantial 
differences in the EWs measured with \texttt{splot} and \texttt{iSpec}. For the blended lines we used the \texttt{deblend} task 
to disentangle absorption lines from
adjacent spectral features.  We gathered $u$ 
photometry and their uncertainties from SDSS DR13 \citep{albareti2017}, 
$G$ photometry and its uncertainty from Gaia DR2
    \citep{gaia2016,gaia2018,arenou2018,evans2018,hambly2018,riello2018}, 
$J$, $H$, and $Ks$ from 2MASS, and $W1$ and $W2$ from WISE.
We use the Gaia DR3 parallax and its uncertainty
\citep{gaiaDR3,fabricius2021,lindegren2021a,lindegren2021b,torra2021}
as well as an extinction $A_V$ inference based on three-dimensional
(3D) maps of extinction in the solar neighborhood from the STructuring
by Inversion the Local Interstellar Medium (Stilism) program
\citep{lallement2014,lallement2018,capitanio2017}. 
We assume \citet{asplund2021} solar abundances. To derive the stellar parameters, we use the  \texttt{isochrones} package by \citet{Morton} to fit the MESA Isochrones and Stellar Tracks \citep[MIST, e.g.][]{Choi,Paxton2011,Paxton2013,Paxton2015} to the photospheric parameters as well as the multiwavelength photometry, parallax, and
extinction using the nested sampling code \texttt{MultiNest} \citep{Feroz2009,Feroz2013}. We present the 
stellar parameters in Table \ref{tab:stellar_param_table}. We would like to remind the readers that the stellar parameters derived from isochrone models are often subject to substantial systematic uncertainties between different isochrone model \citep[e.g. 4\% in stellar radius,][]{Tayar}.  These are not explicitly included in the reported values here. As an additional check we inferred $T_{\text{eff}}$ of TOI-2018 using
the \texttt{colte} code\footnote{\url{https://github.com/casaluca/colte}}
\citep{casagrande2021} that estimates $T_{\text{eff}}$ using a combination
of color--$T_{\text{eff}}$ relations obtained by implementing the InfraRed
Flux Method for Gaia and 2MASS photometry.  As required by \texttt{colte},
we used Gaia DR3 $G$, $G_{\text{BP}}$, and $G_{\text{RP}}$ plus 2MASS
$J$, $H$, and $K_s$ photometry as input.  We find a \texttt{colte}-based
$T_{\text{eff}} = 4160 \pm 81$ K, consistent with our isochrone-inferred 
effective temperature. We could not measure the rotational broadening $v$sin$i$ of the host star given the resolution of our HIRES spectrum. The $v$sin$i$ is likely smaller than 2\,km~s$^{-1}$ (this is consistent with the $23.5$-day rotation period we determined in Section \ref{sec:transit}).

\subsection{Chemical Abundances}

To infer the elemental abundances, we first measured
the equivalent widths of atomic absorption lines of 
\ion{Na}{1}, \ion{Mg}{1}, \ion{Al}{1}, \ion{Si}{1},
\ion{K}{1}, \ion{Ca}{1}, \ion{Sc}{2}, \ion{Ti}{1}, \ion{Ti}{2},
\ion{V}{1}, \ion{Cr}{1}, \ion{Fe}{1}, \ion{Fe}{2}, \ion{Ni}{1}, 
\ion{Co}{1}, \ion{Y}{2}, and \ion{Ba}{2} in our continuum-normalized 
spectrum by fitting Gaussian 
profiles with \texttt{iSpec}. We avoid 
blended lines, and only kept lines with EWs smaller than 170 m\AA. 
We assume \citet{asplund2021} solar abundances and local thermodynamic
equilibrium (LTE) and used the 1D plane-parallel solar-composition
MARCS model atmospheres \citep{Gustafsson} and the 2019 version of \texttt{MOOG} \citep{Sneden1973,Sneden} to
infer elemental abundances based on each equivalent width measurement.
We report our adopted atomic data, equivalent width measurements, and
individual line-based abundance inferences in Table \ref{measured_ews}. We report our abundance inferences
in three common systems: $A(\text{X})$, $[\text{X/H}]$,
and $[\text{X/Fe}]$.  The abundance $A(\text{X})$ is defined
$A(\text{X})=\log{N_{\text{X}}/N_{\text{H}}} + 12$, the abundance
ratio $[\text{X/H}]$ is defined as $[\text{X/H}] = A(\text{X}) -
A(\text{X})_{\odot}$, and the abundance ratio $[\text{X/Fe}]$ is
defined as $[\text{X/Fe}] = [\text{X/H}] - [\text{Fe/H}]$.  We define
the uncertainty in the abundance ratio $\sigma_{[\text{X/H}]}$ as the
standard deviation of the individual line-based abundance inferences
$\sigma_{[\text{X/H}]}'$ divided by $\sqrt{n_{\text{X}}}$ where n is the number of lines used.  We define
the uncertainty $\sigma_{[\text{X/Fe}]}$ as the square root of the sum
of squares of $\sigma_{[\text{X/H}]}$ and $\sigma_{[\text{Fe/H}]}$. The results are reported in Tab. \ref{elem_abundances}.

\begin{deluxetable*}{cccccc}
\tablecaption{Atomic data, Equivalent Widths and line Abundances. Full version online.\label{measured_ews}}
\tablewidth{0pt}
\tablehead{
\colhead{Wavelength} & \colhead{Species} &
\colhead{Excitation Potential} & \colhead{log($gf$)} &
\colhead{EW} & \colhead{$\log_\epsilon(\rm{X})$} \\ 
\colhead{(\AA)} &  & \colhead{(eV)} & & (m\AA) & }
\startdata
$4751.822$ & \ion{NaI}{1} & $2.104$ & $-2.078$ & $41.30$ & $5.824$\\ 
$6154.225$ & \ion{NaI}{1} & $2.102$ & $-1.547$ & $82.90$ & $5.721$\\ 
$6160.747$ & \ion{NaI}{1} & $2.104$ & $-1.246$ & $95.30$ & $5.528$\\ 
$4730.029$ & \ion{MgI}{1} & $4.346$ & $-2.347$ & $110.30$ & $7.475$\\ 
$5711.088$ & \ion{MgI}{1} & $4.346$ & $-1.724$ & $116.30$ & $7.004$\\ 
$6318.717$ & \ion{MgI}{1} & $5.108$ & $-2.103$ & $37.60$ & $7.498$\\ 
$4512.268$ & \ion{CaI}{1} & $2.526$ & $-1.900$ & $63.50$ & $5.617$\\ 
$5260.387$ & \ion{CaI}{1} & $2.521$ & $-1.719$ & $71.70$ & $5.650$\\ 
$5512.980$ & \ion{CaI}{1} & $2.933$ & $-0.464$ & $167.10$ & $5.462$\\ 
$5867.562$ & \ion{CaI}{1} & $2.933$ & $-1.570$ & $64.40$ & $5.771$\\ 
$6166.439$ & \ion{CaI}{1} & $2.521$ & $-1.142$ & $126.30$ & $5.352$
\enddata
\tablecomments{This table is published in its entirety in the machine-readable format.  A portion is shown here for guidance regarding its form and content.} 
\end{deluxetable*}

\begin{deluxetable}{lcccccc}
\tablecaption{Elemental Abundances}\label{elem_abundances}
\tablewidth{0pt}
\tablehead{
\colhead{Species} &
\colhead{$A(\text{X})$} &
\colhead{[X/H]} &
\colhead{$\sigma_{\text{[X/H]}}$} & 
\colhead{[X/Fe]} &
\colhead{$\sigma_{[\text{X/Fe}]}$} &
\colhead{$n$}}
\startdata
\multicolumn{6}{l}{\textbf{LTE abundances}} \\
\ion{Na}{1} & $5.691$ & $-0.529$ & $0.123$ & $0.051$ & $0.089$ & $3$ \\ 
\ion{Mg}{1} & $7.326$ & $-0.224$ & $0.228$ & $0.356$ & $0.169$ & $3$ \\ 
\ion{Al}{1} & $5.796$ & $-0.634$ & $0.031$ & $-0.054$ & $0.032$ & $2$ \\ 
\ion{Si}{1} & $7.230$ & $-0.280$ & $0.246$ & $0.300$ & $0.260$ & $2$ \\ 
\ion{Ca}{1} & $5.570$ & $-0.730$ & $0.147$ & $-0.150$ & $0.077$ & $5$ \\ 
\ion{Sc}{1} & $2.622$ & $-0.518$ & $0.207$ & $0.062$ & $0.151$ & $3$ \\ 
\ion{Ti}{2} & $4.624$ & $-0.346$ & $0.194$ & $0.234$ & $0.099$ & $7$ \\ 
\ion{V}{1} & $3.543$ & $-0.357$ & $0.281$ & $0.223$ & $0.120$ & $7$ \\ 
\ion{Cr}{2} & $4.951$ & $-0.669$ & $0.000$ & $-0.089$ & $0.120$ & $1$ \\ 
\ion{Fe}{1} & $6.880$ & $-0.580$ & $0.171$ & $\cdots$ & $\cdots$ & $70$ \\ 
\ion{Fe}{2} & $6.963$ & $-0.497$ & $0.082$ & $\cdots$ & $\cdots$ & $4$ \\ 
\ion{Co}{1} & $4.724$ & $-0.216$ & $0.185$ & $0.364$ & $0.087$ & $6$ \\ 
\ion{Ni}{1} & $5.947$ & $-0.253$ & $0.307$ & $0.327$ &t $0.105$ & $13$ \\ 
\ion{Cu}{1} & $4.035$ & $-0.144$ & $0.062$ & $0.436$ & $0.059$ & $4$ \\ 
\ion{Y}{2} & $1.260$ & $-0.950$ & $0.000$ & $-0.370$ & $0.033$ & $1$ \\ 
\ion{Ba}{2} & $1.519$ & $-0.751$ & $0.001$ & $-0.171$ & $0.025$ & $2$  
\enddata
\end{deluxetable}

\subsection{Age, SED, Thick Disk membership}
\begin{figure*}
\center
\includegraphics[width = 1.\columnwidth]{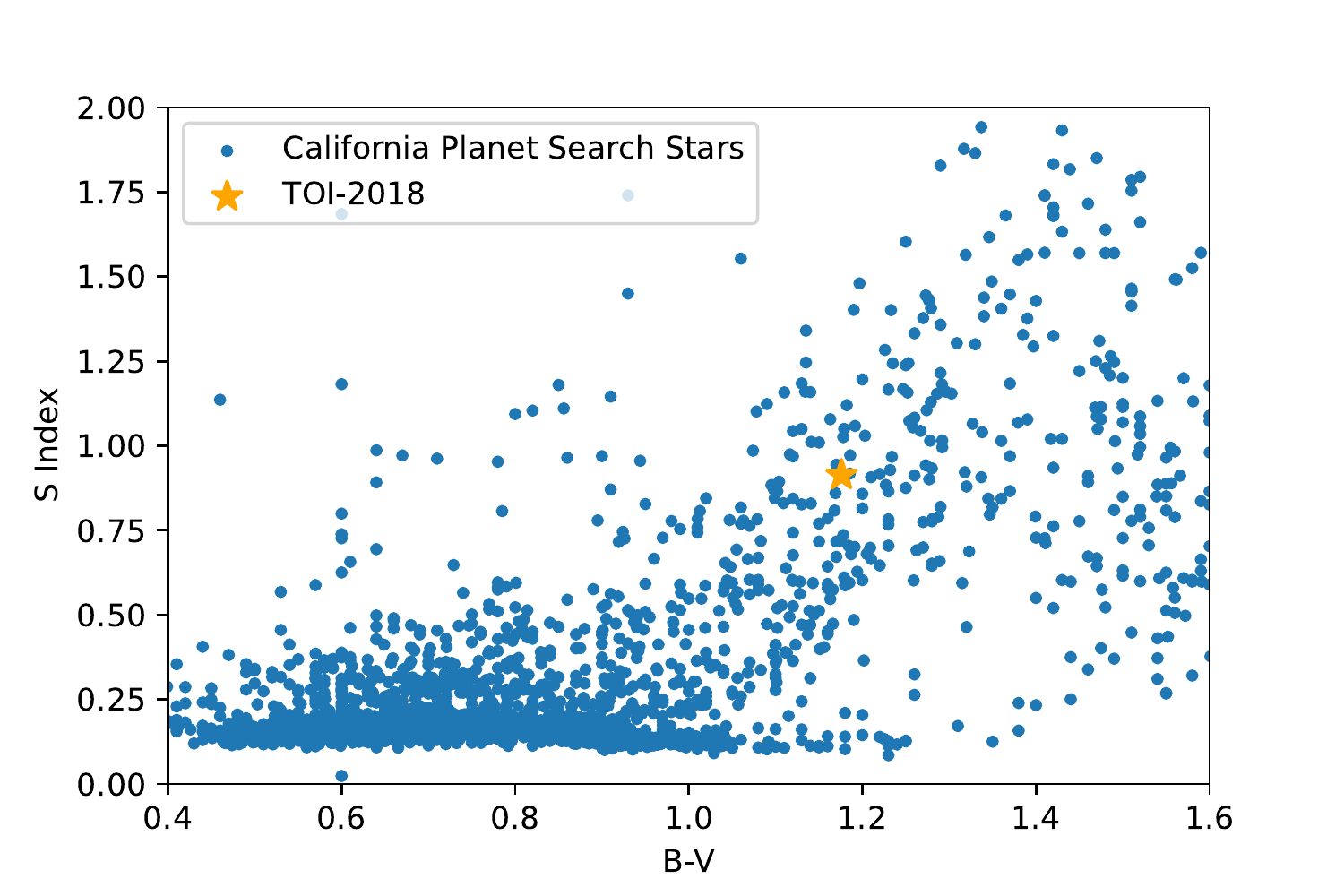}
\includegraphics[width = 1.\columnwidth]{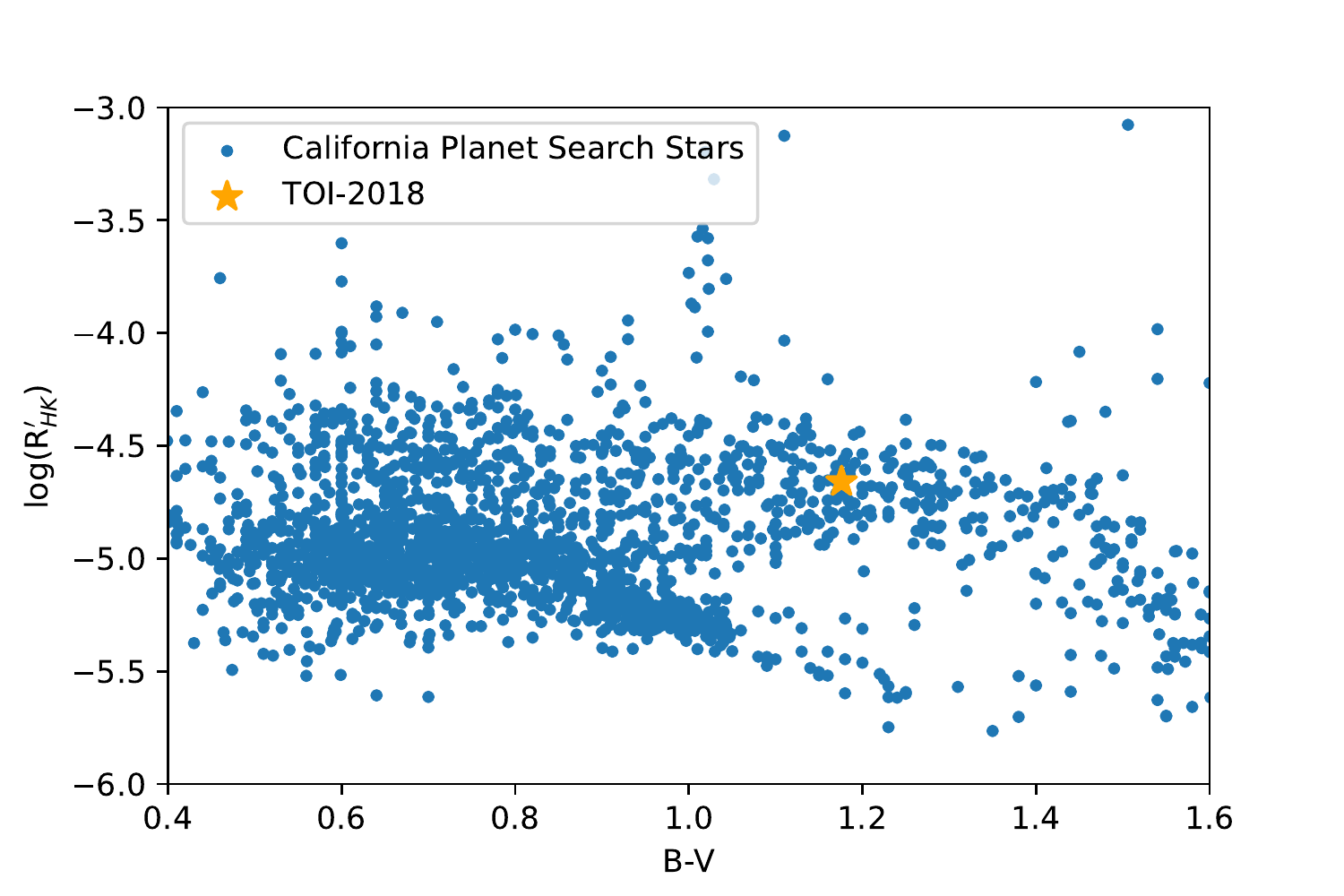}
\caption{The chromospheric activity of TOI-2018 (orange star) in comparison with other stars in the California Planet Search sample \citep{Isaacson}. Both $S_{HK}$ and log$R_{\rm HK}^\prime$ are close to 50\% percentile of stars with similar B-V color. Unfortunately, the star is too cool for applying previously calibrated age-activity relations \citep[e.g.][]{Mamajek}.}
\label{fig:s-index}
\end{figure*}

To check on the system age, we measured a stellar rotation period of $23.5\pm1.0$ days in the WASP light curve of TOI-2018 using the Lomb-Scargle periodogram \citep{Lomb1976,Scargle1982}; auto-correlation function \citep{McQuillan2014} gives a consistent result. The rotation period translates to a gyrochronological age of 1.6$\pm$0.1 Gyr according to the scaling relation of \citet{Mamajek}. If we use the more up-to-date empirical relations of \citet{Bouma_2023}, TOI-2018's rotation period indicates an age of $2.4\pm0.2$Gyr. We also analyzed the chromospheric activity as seen in the Ca II H\&K lines of our HIRES spectrum. We found activity indicator $S_{HK}$=$0.92 \pm 0.03$ and log$R_{\rm HK}^\prime$=-4.75 $\pm$ 0.04 using the method of \citet{Isaacson}. The activity level of TOI-2018 is at about 50\% percentile (see Fig. \ref{fig:s-index}) of stars with similar B-V color (within 0.1 in B-V) observed by the California Planet Search \citep{Howard}. In addition, we looked for Lithium absorption in our HIRES spectrum of TOI-2018. We could not detect a Lithium feature that is statistically significant above the nearby continuum. We note that both the strength of Lithium absorption and log$R_{\rm HK}^\prime$ are likely correlated with host star metallicity [Fe/H]. The existing samples are dominated by solar-metallicity stars.  If TOI-2018 is indeed a thick-disk star as we will discuss shortly, one might expect it to be old ($>$8-9 Gyr). A previous work by \citet{Martig}, however, reported a curious sample of young, $\alpha$-enhanced stars in the solar neighborhood. Given the quality and discrepancy of the various age indicators, we are unable to provide a precise age constraint on TOI-2018 as is often the case for late-type stars. 


We further examined the kinematics of TOI-2018. The proper motion of TOI-2018 does not fit any known comoving associations reported in {\tt Banyan-$\Sigma$} \citep{Gagne} and in \citet{Bouma_2022}. We also computed the Galactic UVW velocity of TOI-2018 (U,V,W = -59.6,  11.4, -8.9 km~s$^{-1}$.). Using the framework of \citet{Bensby}, TOI-2018 has a 3.1\% chance of being in the thick disk based on its kinematics alone. 

\begin{figure}
\hspace{-0.5in}
\includegraphics[width = 1.2\columnwidth]{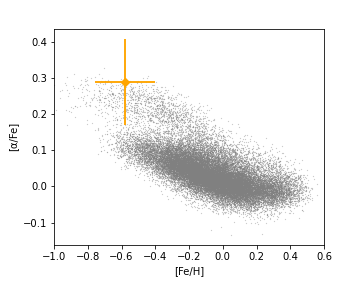}
\caption{TOI-2018 is $\alpha$-element enhanced and [$\alpha$/Fe] depleted which is suggestive of a thick disk star. However, the kinematics (UVW) of TOI-2018 only gives a 3.1\% chance of being a thick disk star. The thick-disk membership of the system needs further confirmation. The gray points are [$\alpha$/Fe] and [Fe/H] for the {\it TESS} Input Catalog \citep{Stassun} as presented in \citet{Carrillo}.}
\label{fig:alpha}
\end{figure}

We also investigated the $\alpha$-element enhancement of TOI-2018. Using Mg, Si, and Ti abundances as a proxy for the $\alpha$ elements (we excluded Ca due to its association with stellar activity), we obtained a [$\alpha$/Fe] = 0.29$\pm0.12$. In Fig. \ref{fig:alpha}, we plot the [$\alpha$/Fe] against [Fe/H] for TOI-2018 and a cross-match between the GALAH survey \citep[e.g.][]{Buder} and the {\it TESS} Input Catalog \citep{Stassun} as presented in \citet{Carrillo}. The  $\alpha$-element enhancement of TOI-2018 does favor a thick disk membership. However, this claim needs to be further confirmed with more precise [$\alpha$/Fe] and kinematic constraints.

\begin{figure}
\hspace{-0.5in}
\includegraphics[width = 1.2\columnwidth]{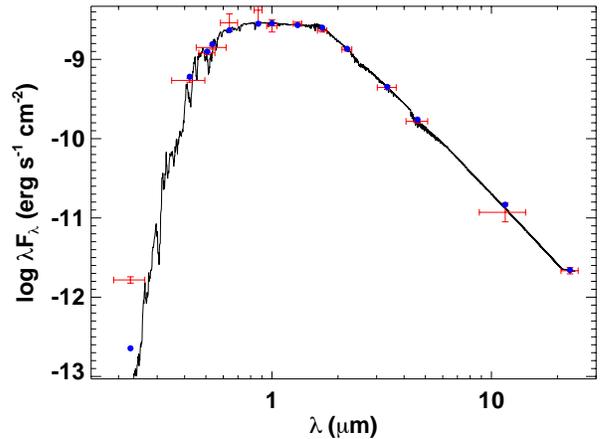}
\caption{The Spectral Energy Distribution (SED) of TOI-2018. The red symbols are the various reported photometric measurements of the system (see text). The horizontal errorbars indicate the effective widths of the passbands. The black curve is the best-fit Kurucz atmosphere model \citep{Kurucz} while the blue symbols are the integrated model flux within each passband. No obvious Infrared excess is detected.}
\label{fig:sed}
\end{figure}

We fitted the Spectral Energy Distribution (SED) of TOI-2018 following the method of \cite{Stassun_SED}. We fitted Kurucz stellar atmosphere models \citep{Kurucz} to various photometric bands. Our fit yielded a reduced $\chi^2$ of 1.1. We obtained a stellar mass and radius of $0.59\pm0.04 M_\odot$ and $0.62\pm0.02R_\odot$ which are consistent with our isochronal analysis. 
We did not find any evidence for an infrared excess that may be attributable to a debris disk (Fig. \ref{fig:sed}).

\subsection{High Resolution Imaging}

As part of our standard process for validating transiting exoplanets, and to assess the contamination of bound or unbound companions on the derived planetary radii \citep{ciardi2015}, we observed TOI-2018 with high-resolution imaging. The star was observed with Palomar/PHARO \citep{hayward2001}, Lick/ShARCS \citep{kupke2012,gavel2014,mcgurk2014}, Gemini-N/Alopeke \citep{Howell}, Caucasian Observatory of
Sternberg Astronomical Institute/Speckle Polarimeter \citep{Safonov2017}, and Carlo Alto/AstraLux \citep{hormuth08} thanks to the efforts of the TESS Follow-up Observing Program (TFOP) Working Group.   

We present the Palomar/PHARO result here as an example. All other high-resolution imaging results did not detect any nearby stellar companions, and they are available on the ExoFOP website \footnote{\url{https://exofop.ipac.caltech.edu/tess/}}. The Palomar observations were made with the PHARO instrument \citep{hayward2001} behind the natural guide star AO system P3K \citep{dekany2013} on 2021~Jun~19 UT in a standard 5-point quincunx dither pattern with steps of 5\arcsec\ in the narrow-band $Br-\gamma$ filter $(\lambda_o = 2.1686; \Delta\lambda = 0.0326~\mu$m).  Each dither position was observed three times, offset in position from each other by 0.5\arcsec\ for a total of 15 frames; with an integration time of 1.4 seconds per frame, the total on-source time was 14 seconds. PHARO has a pixel scale of $0.025\arcsec$ per pixel for a total field of view of $\sim25\arcsec$. The science frames were flat-fielded and sky-subtracted.  The reduced science frames were combined into a single combined image with a final resolution of 0.091\arcsec FWHM.

\begin{figure}[h]
    \centering
    \includegraphics[width=0.46\textwidth]{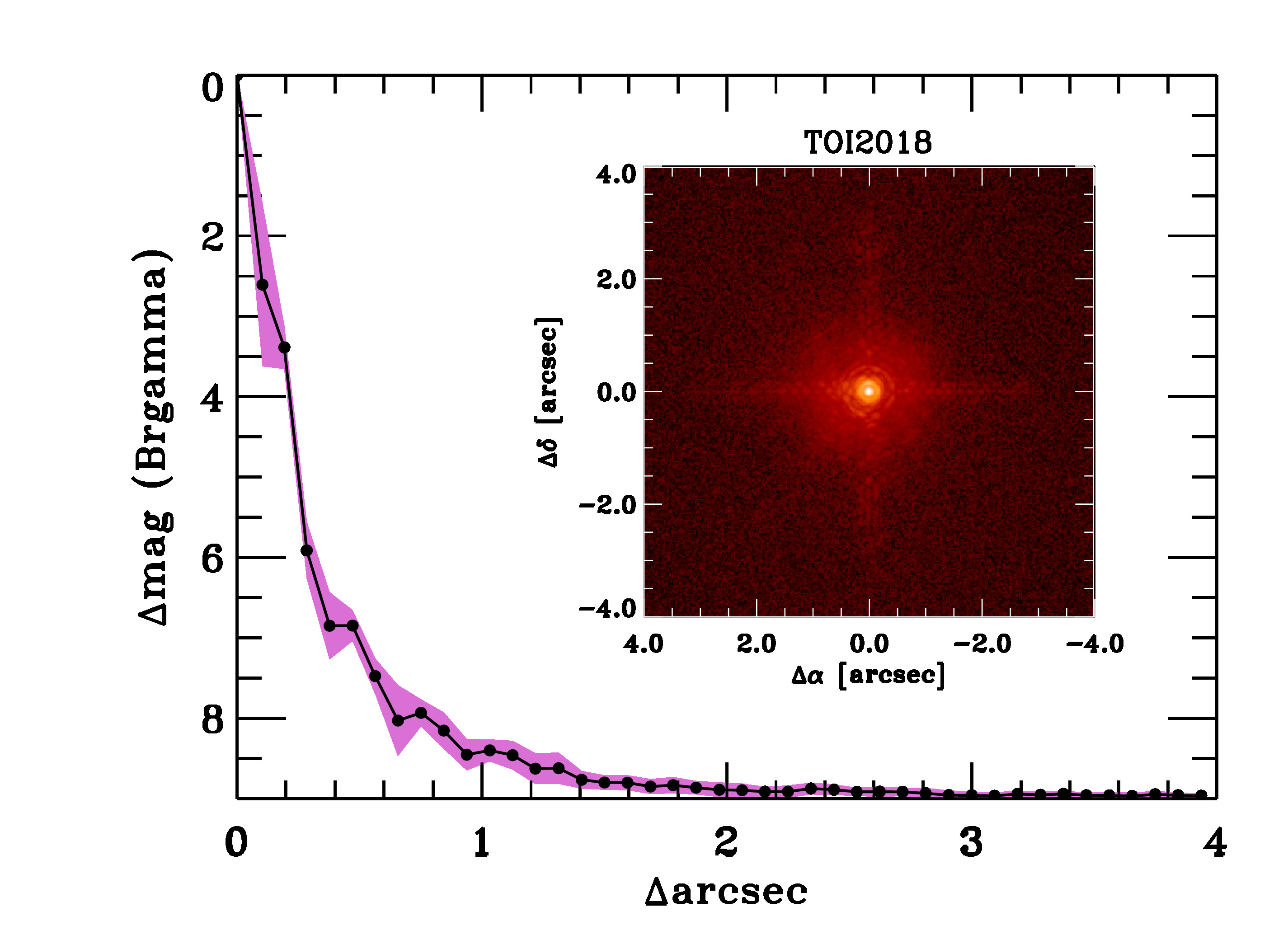}
    \caption{Companion sensitivity for the near-infrared adaptive optics imaging.  The black points represent the 5$\sigma$ limits and are separated in steps of 1 FWHM; the purple represents the azimuthal dispersion (1$\sigma$) of the contrast determinations (see text). The inset image is of the primary target showing no additional close-in companions.}\label{fig:ao_contrast}  
	\vspace{-0.5em}
\end{figure}

To within the limits of the AO observations, no stellar companions were detected. The sensitivities of the final combined AO image were determined by injecting simulated sources azimuthally around the primary target every $20^\circ $ at separations of integer multiples of the central source's FWHM \citep{furlan2017, lund2020}. The brightness of each injected source was scaled until standard aperture photometry detected it with $5\sigma $ significance. The resulting brightness of the injected sources relative to TOI-2018 set the contrast limits at that injection location. The final $5\sigma $ limit at each separation was determined from the average of all of the determined limits at that separation. The uncertainty on the limit was set by the root-mean-square dispersion of the azimuthal slices at a given radial distance (Fig.~\ref{fig:ao_contrast}).

In addition to the high-resolution imaging, we also utilized Gaia to identify any wide stellar companions that may be bound members of the system.  Typically, these stars are already in the {\it TESS} Input Catalog and their flux dilution to the transit has already been accounted for in the transit fits and associated derived parameters.  Based upon similar parallaxes and proper motions \citep[e.g.,][]{mugrauer2020,mugrauer2021,mugrauer2022}, there are no additional widely separated companions identified by Gaia. Additionally, the Gaia DR3 astrometry provides additional information on the possibility of inner companions that may have gone undetected by either Gaia or high-resolution imaging. The Gaia Renormalised Unit Weight Error (RUWE) is a metric similar to a reduced chi-square, where values that are $\lesssim 1.4$  indicate that the Gaia astrometric solution is consistent with the star being single. In contrast, RUWE values $\gtrsim 1.4$ may indicate an astrometric excess noise, possibily caused by the presence of an unseen companion \citep[e.g., ][]{ziegler2020}.  TOI-2018 has a Gaia DR3 RUWE value of 1.23, indicating that the astrometric fits are consistent with the single-star model. 

\section{Photometric Analysis}\label{sec:transit}
\subsection{{\it TESS}  Observations}
TOI-2018  (TIC 357501308) was observed by the {\it TESS} mission \citep{Ricker} in Sectors 24 and0 51. We started with the 2-min cadence light curve produced by the {\it TESS} Science Processing Operations Center \citep[SPOC located at NASA Ames Research Center,][]{jenkinsSPOC2016}. The data was downloaded from the Mikulski Archive for Space
Telescopes website\footnote{\url{https://archive.stsci.edu}} which is available at the following \dataset[DOI]{doi:10.17909/t9-nmc8-f686}. Our subsequent analysis was based on the Presearch Data Conditioning Simple Aperture Photometry \citep[PDCSAP;][]{Stumpe2012,Stumpe2014,Smith2012} version of the light curves, although the Simple Aperture Photometry \citep[SAP, ][]{Twicken2010,Morris2020} version was used to measure the stellar rotation period since it better preserves any long-term stellar variability. We excluded anomalous data points that have non-zero Quality flags. We note that the SPOC pipeline found a difference imaging centroid offset that is only 0.945$\pm$ 2.55''; this is again consistent with a lack of nearby stellar companion for TOI-2018.

\begin{figure*}
\center

\includegraphics[width = 1.5\columnwidth]{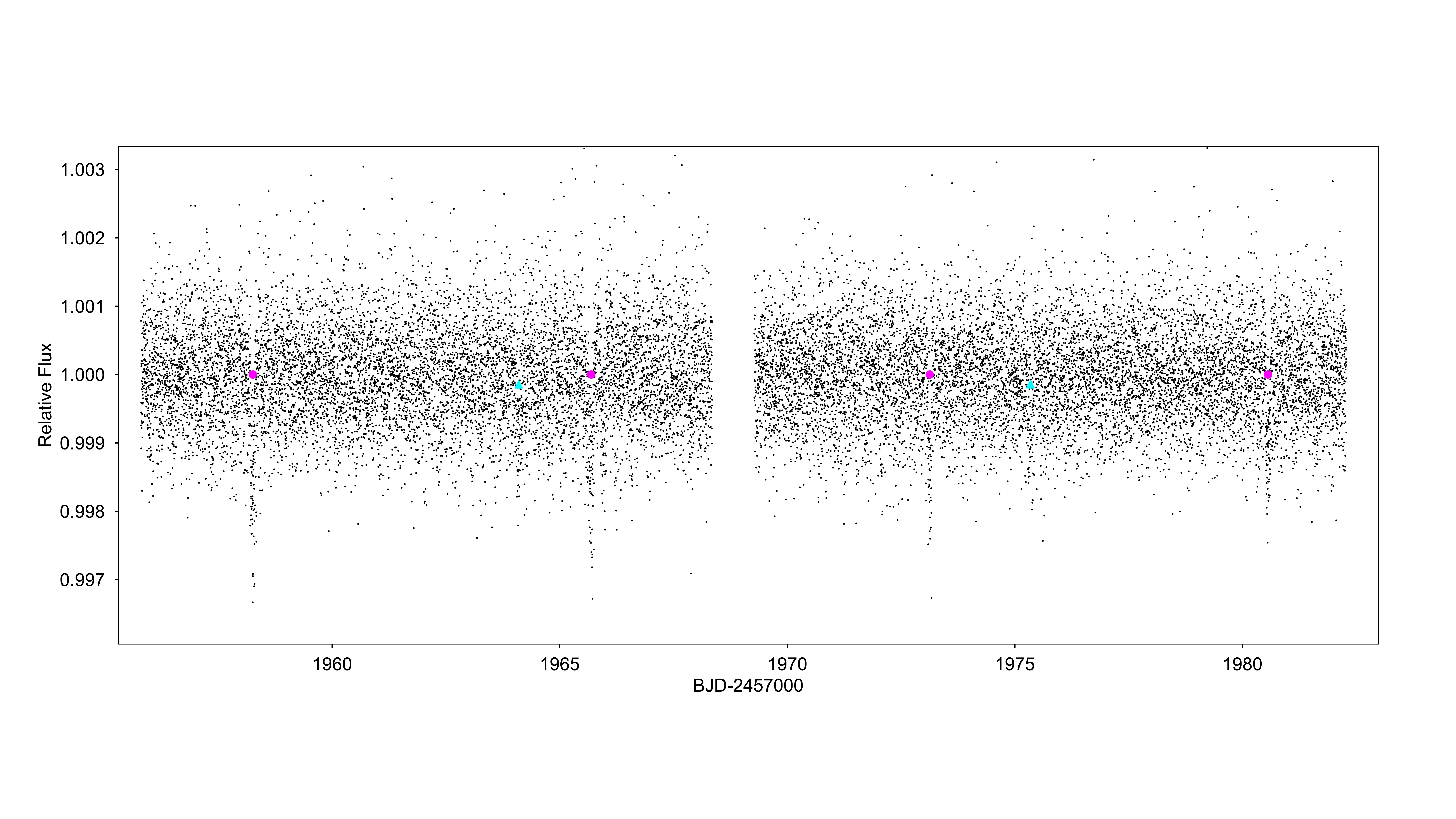}
\includegraphics[width = 1.5\columnwidth]{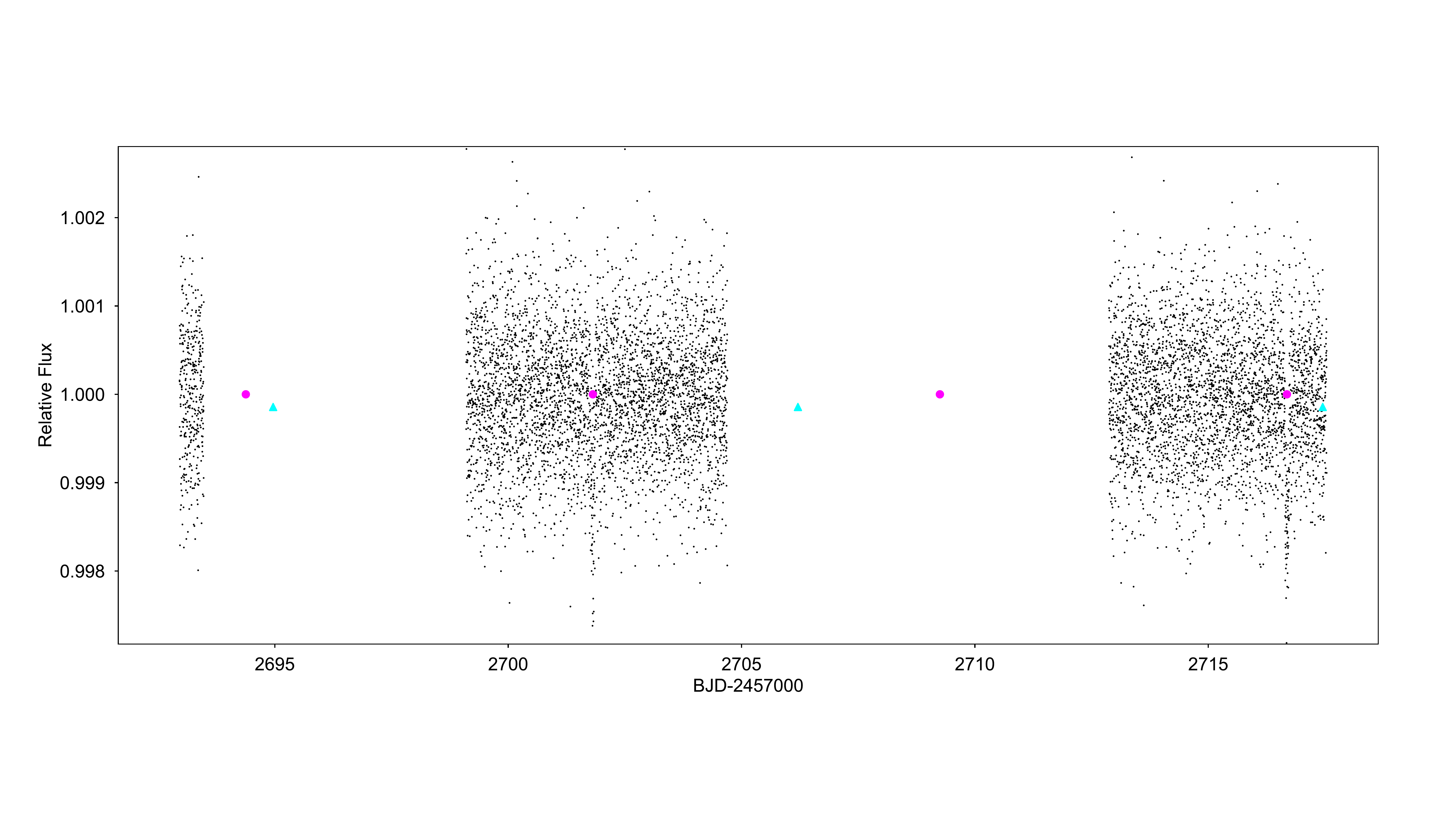}
\caption{The detrended {\it TESS} light curve of TOI-2018. The color symbols indicate the transits of TOI-2018 b (magenta). We also found a second possible transit signal TOI-2018.02 (cyan). The signal is below the typical SNR threshold (5.8 v.s. 7.1) to be counted as a planet candidate by the {\it TESS} team.}
\label{fig:lc}
\end{figure*}

 We removed any stellar activity or instrumental effects by fitting the light curve with a cubic spline in time of a width of 0.75 days. We searched for planetary transit signals in the detrended light curve using the Box-Least-Square algorithm \citep[BLS,][]{Kovac2002}. Our pipeline has previously used for the detection of other K2 and {\it TESS} planets \citep{Dai2017,Dai_1444}. We detected a 7.4-day planet with a signal detection efficiency \citep[as defined by][]{Kovac2002} of 11.7. The same candidate was reported by the {\it TESS} team \citep{Guerrero} as TOI-2018.01. We also detected a second planet candidate at 11.3 day with a signal detection efficiency of 5.8 which is below the SPOC signal detection limit (SNR = 7.1). Given the low SNR of the detection, TOI-2018.02 does not meet the usual threshold for qualifying as a planet candidate. However, the orbital periods of the two planets are close to a 3:2 mean-motion resonance with a small deviation of $(P_{\rm out}/P_{\rm in})/(3/2)-1 \approx 1\%$ which may boost the case for a real planet for TOI-2018.02. We did not find any other significant transit signal in the {\it TESS} light curve. Fig. \ref{fig:lc} shows the detrended light curve and the transits of the two candidate planets.

We used the {\tt Python} package {\tt Batman} \citep{Kreidberg2015} to model the transit light curves of the two planets simultaneously. One of the global parameters is the mean stellar density of the host $\rho = 2.39\pm0.09\rho_\odot$ as derived in Section \ref{sec:stellar_para}. We imposed a Gaussian prior on the mean stellar density to help break the degeneracy in semi-major axis and impact parameter. Two other global parameters are the quadratic limb darkening coefficients using the reparameterization of $q_1$ and $q_2$ suggested by \citet{Kipping}. We adopted a Gaussian prior using the theoretical values from {\tt EXOFAST} \citep{Eastman2013} and a standard deviation of 0.3. Both planets were assumed to have circular orbits, hence the other transit parameters are the orbital period $P_{\text{orb}}$, the time of conjunction $T_{\text{c}}$, the planet-to-star radius ratio $R_{\text{p}}/R_\star$, the scaled orbital distance $a/R_\star$, and the transit impact parameter $b$.

\begin{figure*}
\center
\includegraphics[width = 1.\columnwidth]{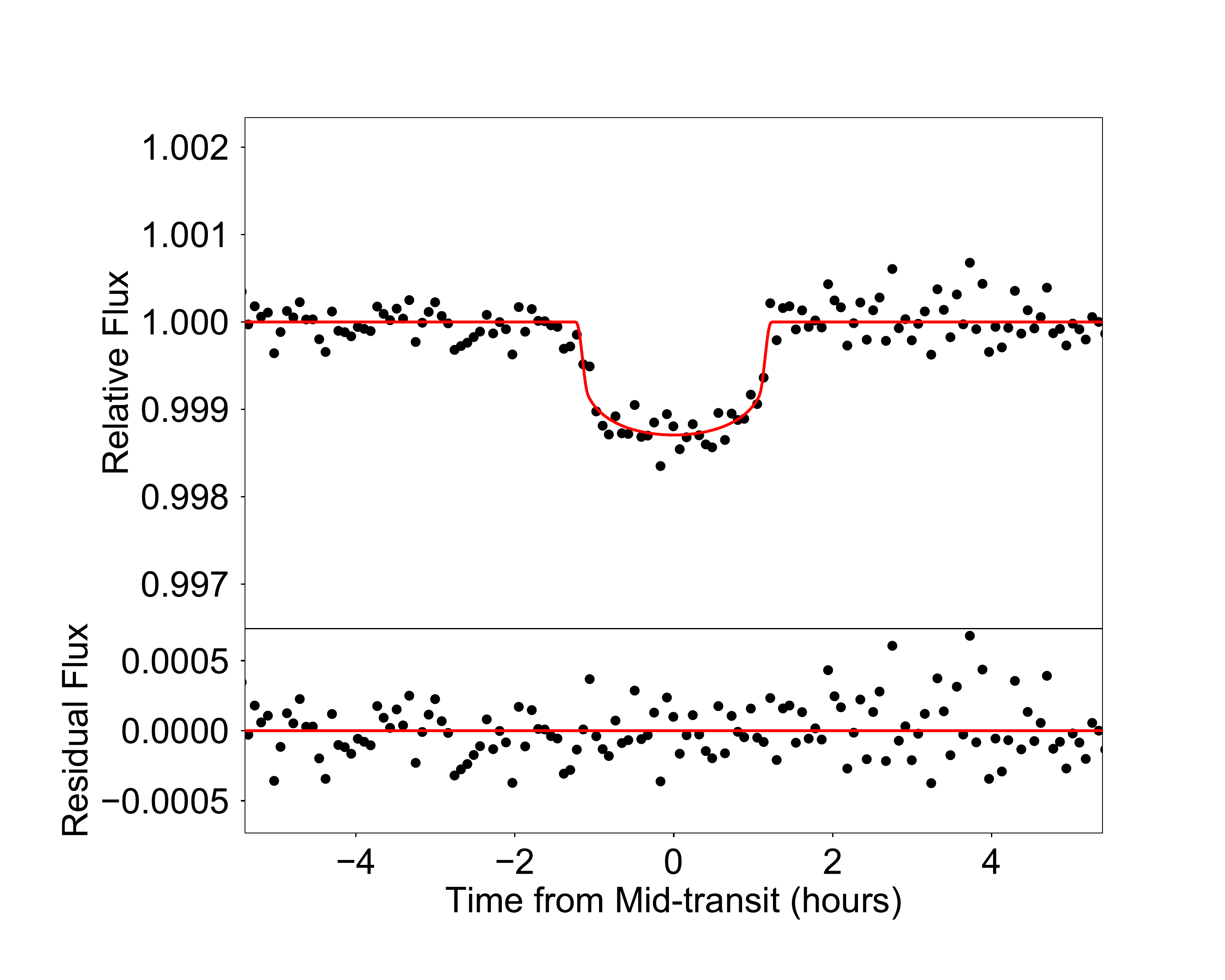}
\includegraphics[width = 1.\columnwidth]{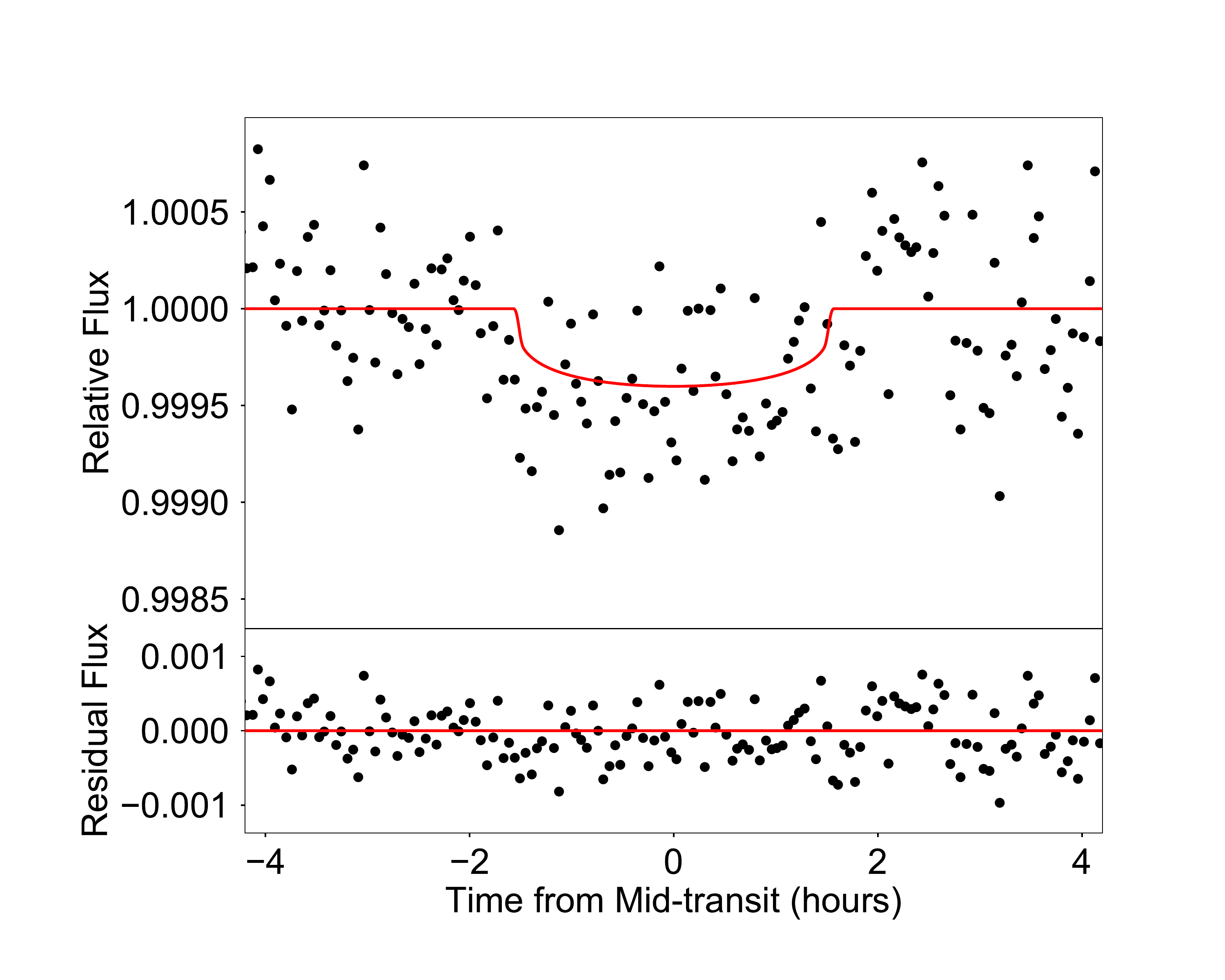}
\caption{The phase-folded and binned transit light curves of TOI-2018 b (left) and TOI-2018.02 (right). The best-fit linear ephemeris model is shown in red. }
\label{fig:transit}
\end{figure*}

We started by assuming that both planet candidates have linear ephemerides (i.e. no transit timing variations). We fitted all transits of each planet with a constant period model. The best-fit model was determined with the {\tt Levenberg-Marquardt} method implemented in {\tt Python} package {\tt lmfit} \citep{LM}.
This best-fit model was used as a template transit to fit the mid-transit times of each individual transit. During the fit of individual transits, we varied only the mid-transit time and three parameters of a quadratic function of time that describes any residual long-term variations. A total of 6 and 3 transits were observed for the two planet candidates. We were not able to detect a statistically significant transit timing variation trend for either planet. We note that the transits of TOI-2018.02 in Sector 51 were either located in data gaps or near the end of the {\it TESS} observation (Fig. \ref{fig:lc}). TOI-2018.02 could not be recovered using Sector 51 data alone. Our ephemeris of TOI-2018.02 is based on a joint fit using all sectors from {\it TESS}, the result has substantial uncertainty (Tab. \ref{tab:planet_para}) due to the ambiguity of the transit time in Sector 51. {\it TESS} will observe this system again in Sector 77 and 78; those data will be instrumental in confirming TOI-2018.02 and for detecting transit timing variations. We carried out a Monte Carlo Markov Chain analysis using the {\tt emcee} package \citep{emcee}. We initialized 128 walkers near the best-fit model from  {\tt lmfit}. We ran the MCMC for 50000 links which is more than two orders of magnitude longer than the typical autocorrelation function ($\lesssim300$ links). The resultant posterior distribution is summarized in Table \ref{tab:planet_para}. Fig.~\ref{fig:transit} shows the phase folded and binned light curves for each planet candidate as well as the best-fit transit model.

\subsection{MuSCAT2 Observations}
We observed two egresses of TOI-2018 b with the multi-band imager MuSCAT2 \citep{Narita2019} mounted on the 1.5 m Telescopio Carlos S\'{a}nchez (TCS) at Teide Observatory, Spain. We obtained simultaneous $g'$, $r'$, $i'$, and $z_s$ photometry on the nights of 15 June 2022 and 17 March 2023. We performed basic data reduction (dark and flat correction), aperture photometry, and transit model fit including systematics with the MuSCAT2 pipeline \citep{Parviainen2019}. On both nights, we detected the egress of TOI-2018 b; the transit did not present any significant transit depth variations across the four MuSCAT2 bands. The transit times align well with those predicted from the {\it TESS} light curve with no apparent transit timing variations.  A joint fit of transit times from {\it TESS} and {\it MuSCAT2} refined the transit ephemeris of TOI-2018 b: $P_{\rm orb}=$7.435569 $\pm$ 0.000081 days; $T_c$ (BJD-2457000) = 1958.25782$\pm$ 0.00058. The MuSCAT2 data is available on the ExoFOP website.

\subsection{WASP Observations}
TOI-2018 was observed by the WASP survey \citep{WASP} from UT May 3 2004 to Jun 28 2007. The data is available at the following link \footnote{\url{https://exoplanetarchive.ipac.caltech.edu/docs/SuperWASPMission.html}}. We could not recover the transit signal of either TOI-2018 b and TOI-2018.02 in the WASP light curve (non-detection is expected given WASP light curve quality). However, the much longer observational baseline of WASP  provides a better constraint on the stellar rotation period than the {\it TESS} light curve. 

\section{Radial Velocity Analysis}
We acquired a total of 38 high-resolution spectra of TOI-2018 on the Keck/HIRES \citep{HIRES} from UT Jun 18 2011 to Aug 30 2021. These spectra were obtained with iodine cell in the path of light. The iodine cell served as the reference for our wavelength solution and the line spread function. The exposure time is typically 300-600 seconds after which we obtained a median SNR of 140 per reduced pixel near 5500 Å. We extracted the radial velocity using our forward-modeling Doppler code described in \citet{Howard}. The estimated radial velocity uncertainty is about 1.2 m~s$^{-1}$. The extracted radial velocities and stellar activity indices are shown in Table \ref{tab:rv}.

We assumed that the planets are on circular orbits. We also experimented with non-zero eccentricities. However, the data at hand is not sufficiently constraining. The posterior samples prefer circular models with a $\Delta$BIC>10. We simplified our analysis by focusing on circular orbits only. With circular orbits, the radial velocity signals are hence described by the orbital period $P_{\rm orb}$, the time of inferior conjunction $T_c$,  and the RV semi-amplitude $K$.  We also included an RV offset $\gamma$, a linear RV trend $\Dot{\gamma}$, and a jitter term $\sigma_{jit}$ to account for any residual astrophysical or instrumental radial velocity uncertainties.  We imposed Gaussian priors on $P_{\rm orb}$ and $T_c$ using the posterior distribution obtained from transit analysis. We imposed log-uniform priors on the RV semi-amplitude $K$ and the jitter $\sigma_{\text{jit}}$. We imposed uniform priors on the RV offset $\gamma$ and linear trend $\Dot\gamma$. To model the influence of stellar activity contamination in the RV dataset, we employed a Gaussian Process (GP) model \citep[e.g. ][]{Haywood,Grunblatt2015,Dai2017} with a quasi-periodic kernel:

\begin{equation}
\label{covar}
\begin{split}
C_{i,j} = h^2 \exp{\left[-\frac{(t_i-t_j)^2}{2\tau^2}-\Gamma \sin^2{\frac{\pi(t_i-t_j)}{T}}\right]} \\ 
+\left[\sigma_i^2+\sigma_{\text{jit}}^2\right]\delta_{i,j}
\end{split}
\end{equation}
where $t_i$ is the time of individual RV measurements; $C_{i,j}$ is the covariance matrix; $\delta_{i,j}$ is the Kronecker delta function;
$h$ is the amplitude of the covariance; $\tau$ is the correlation timescale; $\Gamma$ quantifies the relative significance between the squared exponential and periodic parts of the kernel; $T$ is the period of the covariance; $\sigma_i$ is the internal RV uncertainty and $\sigma_{jit}$ is the jitter term. 

The corresponding likelihood function is:

\begin{equation}
\label{likelihood}
\log{\mathcal{L}} =  -\frac{N}{2}\log{2\pi}-\frac{1}{2}\log{|\bf{C}|}-\frac{1}{2}\bf{r}^{\text{T}}\bf{C} ^{-\text{1}} \bf{r}
\end{equation}
where $N$ is the total number of RV measurements; and $\bf{r}$ is the residual after subtracting the Keplerian planetary signals from the observed RV variation.

We first trained the GP model on the out-of-transit light curves first. The underlying assumption is that the stellar surface magnetic activity drives both the out-of-transit flux variation in the light curve as well as a spurious quasi-sinusoidal contamination in RV measurement. Since the out-of-transit light curve is measured to higher precision and better sampled. We conditioned all the Gaussian process hyper-parameters on the light curves first before using those hyper-parameters in the radial velocity analysis.

\begin{figure*}
\center
\includegraphics[width = 1.5\columnwidth]{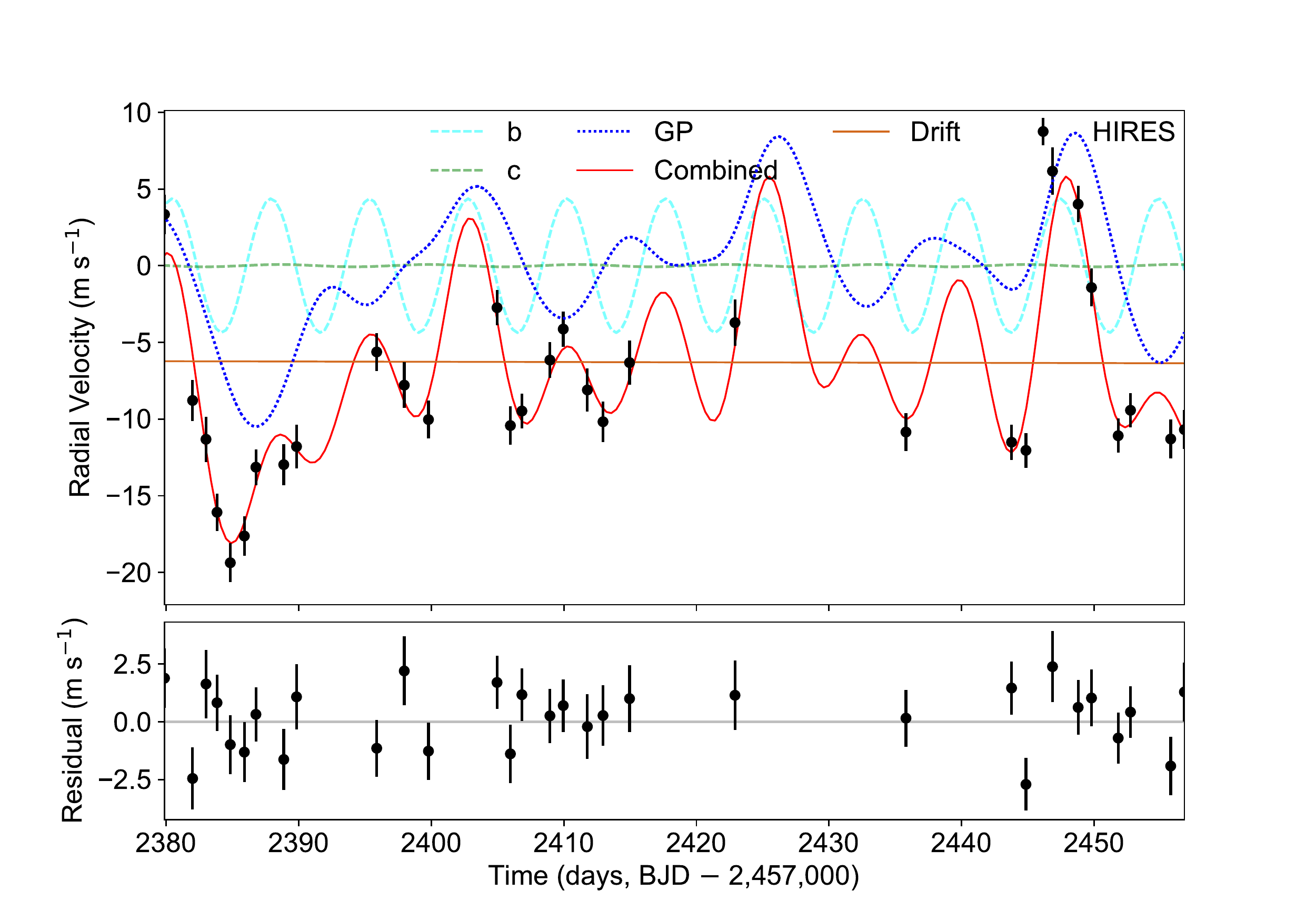}
\includegraphics[width = 1.5\columnwidth]{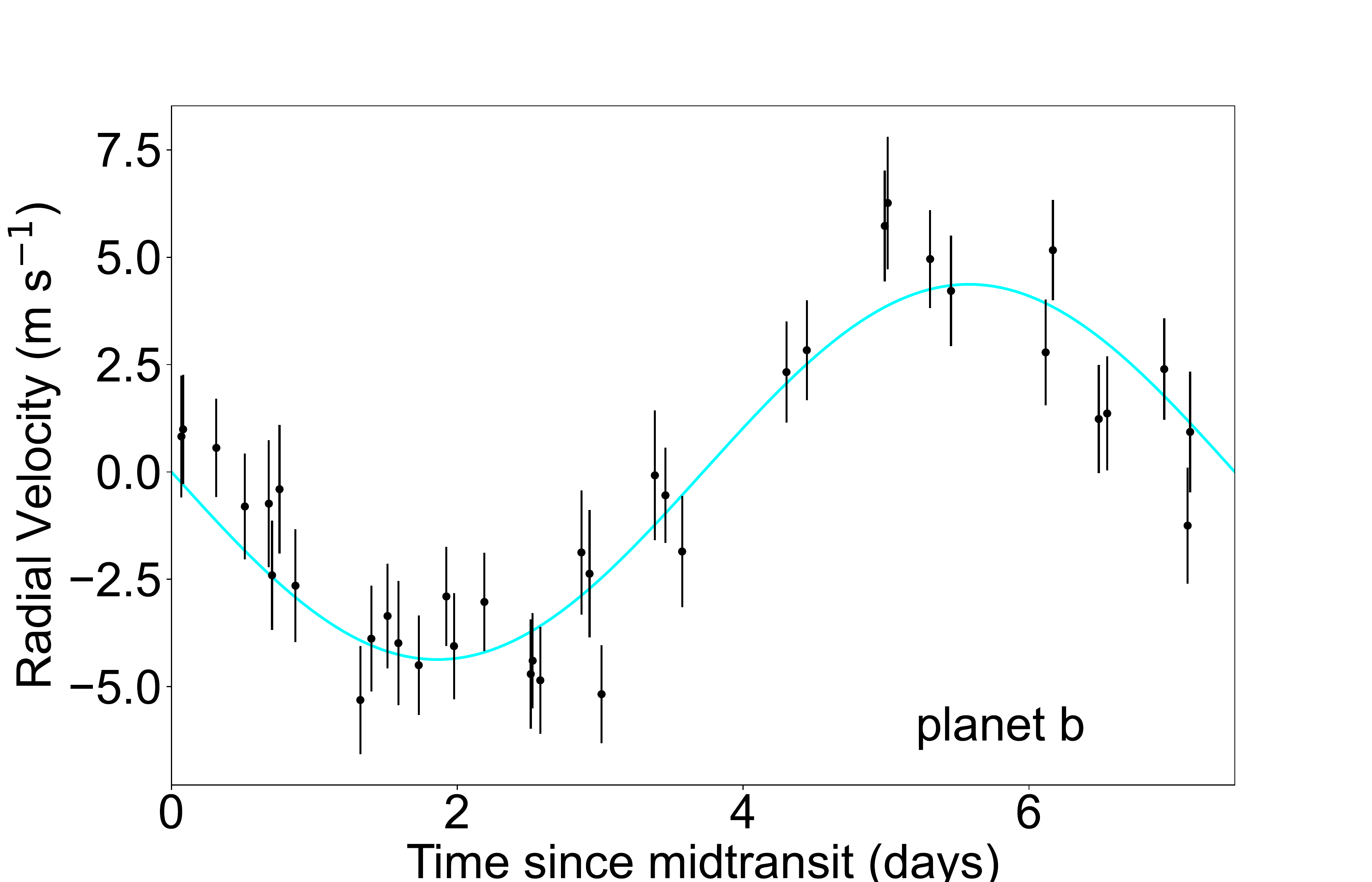}
\caption{Top: The HIRES radial velocity measurement of TOI-2018 (black symbols) and the best-fit RV model (red). The best-fit model only includes contribution from TOI-2018 b (cyan), a radial velocity drift (orange), and a Gaussian Process model for stellar activity (blue).  TOI-2018.02 (green dotted) was not detected in RV. Bottom: the RV measurements phase-folded at the orbital period of TOI-2018 b after removing the contribution from the other effects in the Top panel.}
\label{fig:rv}
\end{figure*}
\begin{figure*}
\center
\includegraphics[width = 1.5\columnwidth]{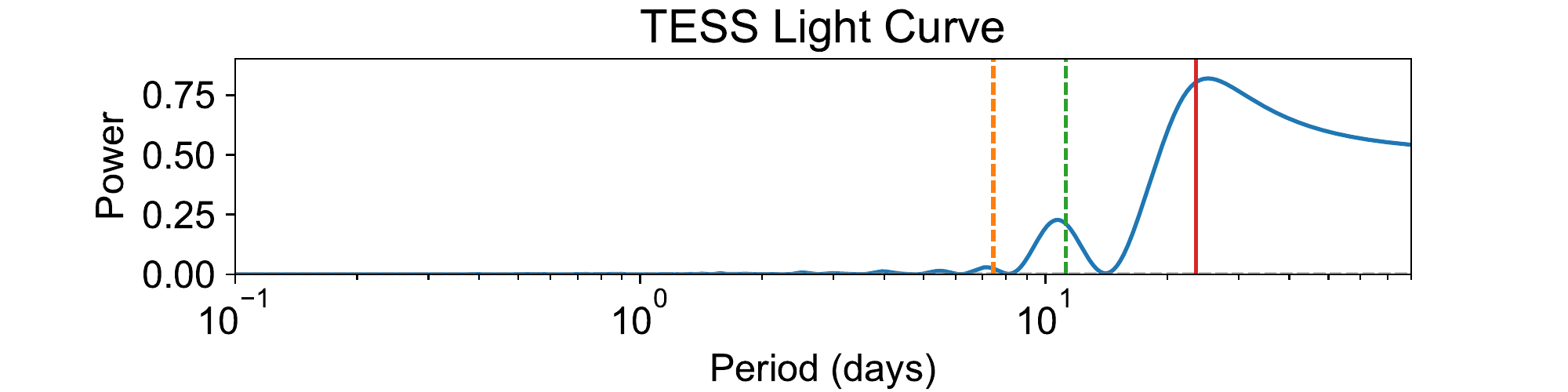}
\includegraphics[width = 1.5\columnwidth]{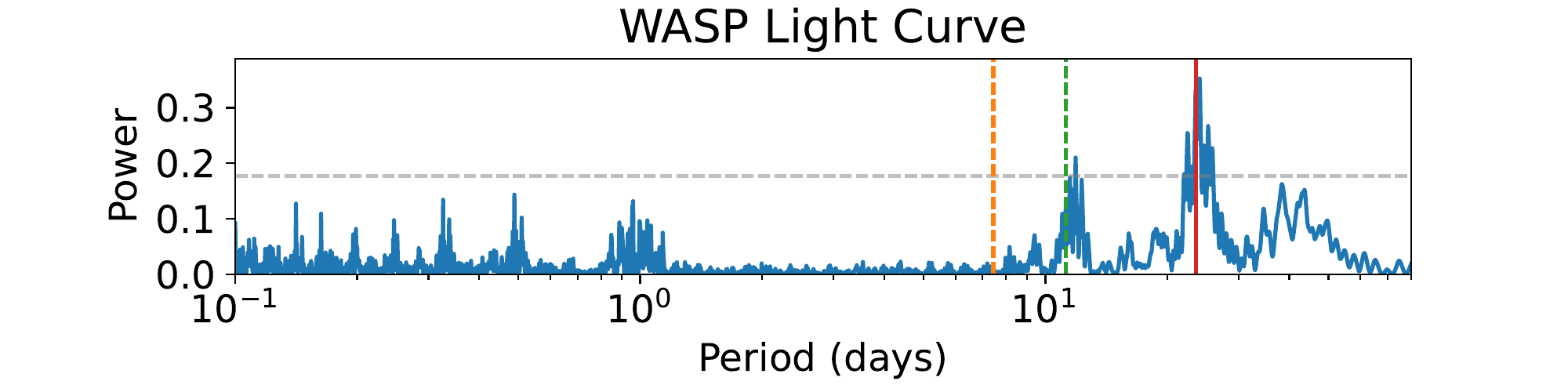}
\includegraphics[width = 1.5\columnwidth]{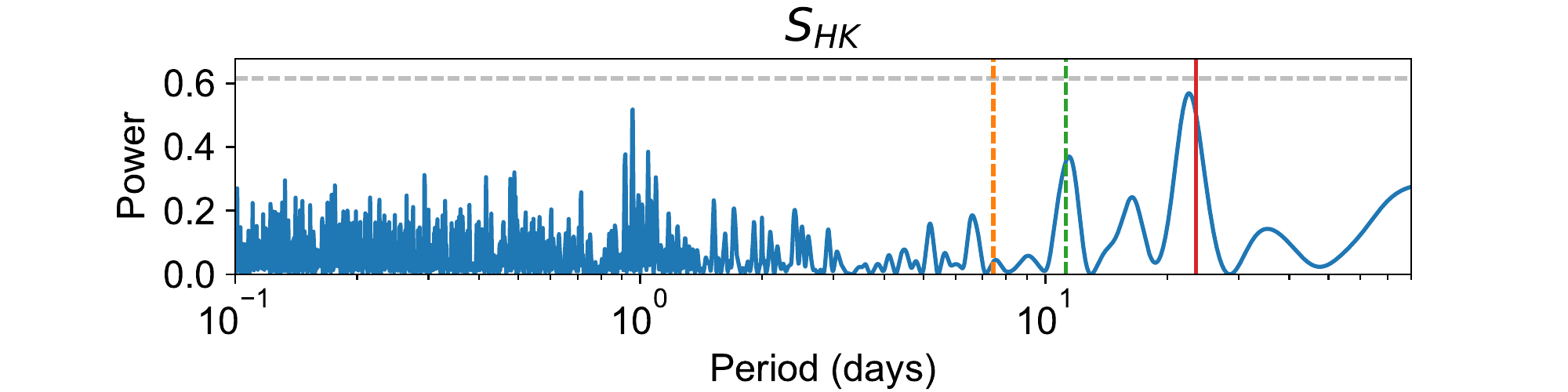}
\includegraphics[width = 1.5\columnwidth]{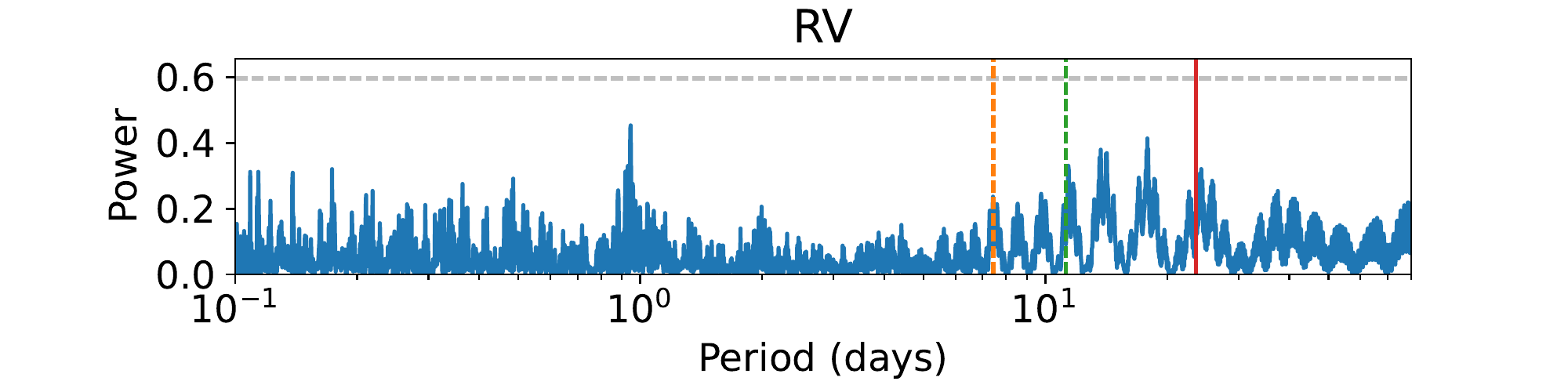}
\includegraphics[width = 1.5\columnwidth]{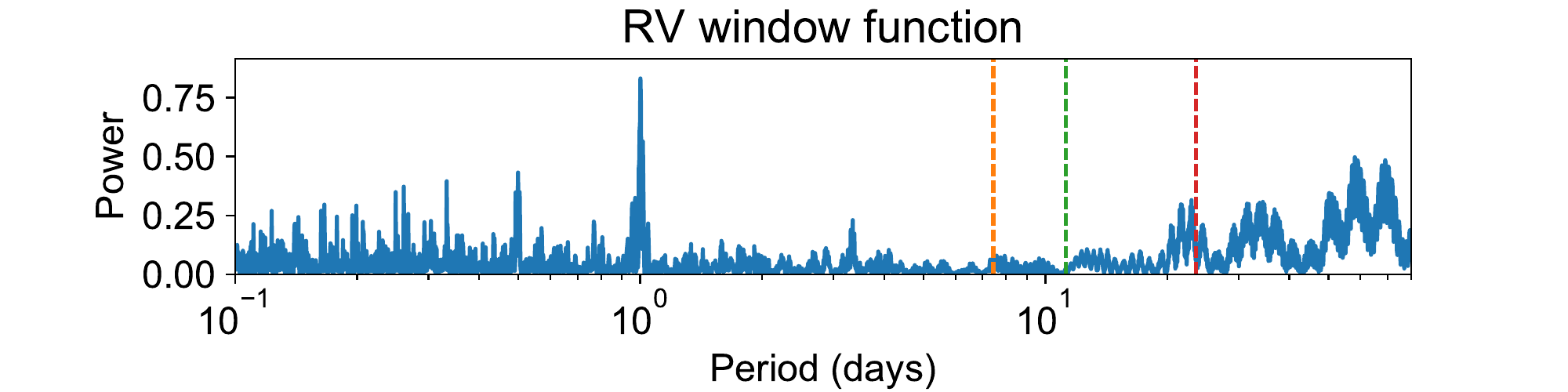}
\includegraphics[width = 1.5\columnwidth]{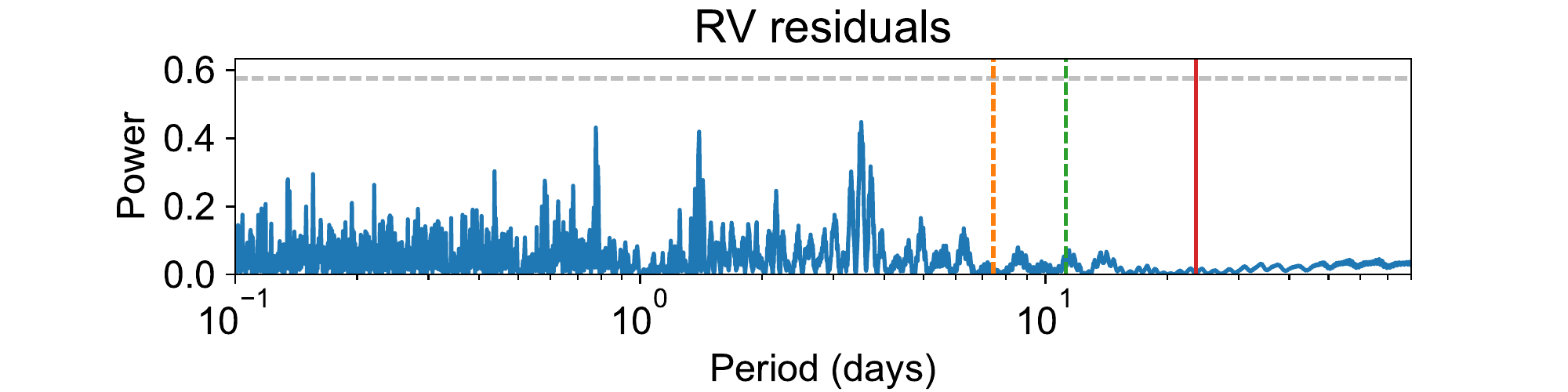}
\caption{The Lomb-Scargle periodograms of the {\it TESS} light curve, the WASP light curve, the HIRES $S_{HK}$ activity indicator, the measured radial velocities, the RV window function, and the RV residuals after subtracting the best-fit model (Fig. \ref{fig:rv}). The vertical lines are respectively the orbital period of TOI-2018 b (orange dashed) and TOI-2018.02 (green dashed), and the rotation period of the host star (red solid). Whenever appropriate, we included horizontal dashed lines to indicate the 1\% false alarm levels.}
\label{fig:periodogram}
\end{figure*}

We experimented with including increasingly more complexity to our radial velocity models. We increased the number of planets included and if a GP model for stellar activity was warranted by the RV data set; and if a linear RV trend $\Dot\gamma$ was required. We selected the best model by examining the Bayesian Information Criterion (BIC) after model optimization with the {\tt Levenberg-Marquardt} method {\tt lmfit} \citep{LM}. The model favored by the current dataset contains only planet b, a long-term RV drift ($\Dot{\gamma}$), and a GP model for the stellar activity. We sampled the posterior distribution of this model using a similar sampling procedure as described in Section \ref{sec:transit} using {\sc emcee}. We performed two separate samplings. The posterior distribution of the various hyperparameters from an MCMC analysis of the WASP light curve was used a prior for a subsequent RV analysis. The underlying assumption is that the light curve is dominated by quasi-periodic flux variations due to the host's stellar activity. We summarize the posterior distribution of RV analysis in Table \ref{tab:planet_para}. The radial velocity variation of planet b is securely detected with more than 4-$\sigma$ significance (Fig. \ref{fig:rv}).  We note that the radial velocity alone was able to independently discover planet b, the ephemeris of the planet from the transit analysis was crucial for recovering the radial velocity signal. A linear RV drift $\Dot{\gamma}$ is marginally detected at -0.0017$\pm$0.0008 m~s$^{-1}$~day$^{-1}$ over the 10-year baseline of our HIRES observation. Unfortunately, the orbital period of TOI-2018.02 is close to the first harmonic of the rotation period (11.3-day v.s. 23.5/2 days, Fig. \ref{fig:periodogram}). We were only able to place an upper limit on the mass of TOI-2018.02 (<3.6$M_\oplus$ or $K<1.5$m~s$^{-1}$ at a 95\% confidence level). Given the RV non-detection and the fact that only three transits were observed by {\it TESS}, we report TOI-2018.02 only as a possible planet candidate.

\begin{deluxetable*}{llcc}
\tablecaption{Model Parameters of TOI-2018\label{tab:planet_para}}
\tablehead{
\colhead{Parameter}  & \colhead{Symbol} &  \colhead{TOI-2018 b} &  \colhead{TOI-2018.02} }
\startdata
{\bf From Transit Modeling}\\
Mean Stellar Density ($\rho_\odot$)& $\rho_\star$  &
 $2.29\pm0.22$ &-\\
Limb Darkening Coefficient & $q1$  &
 $0.36\pm0.22$ &-\\
Limb Darkening Coefficient & $q2$ &
 $0.33\pm0.23$&- \\
Orbital Period (days) & $P_{\rm orb}$  
 &$7.435583\pm0.000022$  
 &$11.244\pm0.025$\\
Time of Conjunction (BJD-2457000) & $t_c$  
 &$1958.2580\pm0.0013$  
 &$1964.110\pm0.020$\\
 Planet/Star Radius Ratio & $R_p/R_\star$  
 & $0.0335\pm0.0010$
 & $0.0228\pm0.0020$\\
Impact Parameter & $b$  
 & $0.546\pm0.070$
 & $0.16\pm0.11$\\
Scaled Semi-major Axis & $a/R_\star$ 
 &$21.12\pm0.69$  
 &$27.82\pm0.90$\\
Transit Duration (hours)& $T_{\rm 14}$
 & $2.360\pm0.090$
 & $3.09\pm0.12$\\
Orbital Inclination (deg)   & $i$   
&$88.52\pm0.22$
 & $89.66\pm0.31$\\
 Orbital Eccentricity    & $e$   
&0 (fixed)
 & 0 (fixed)\\
 \hline
{\bf From Radial Velocity Modeling}\\
Semi-Amplitude (m~s$^{-1}$)& $K$  
 &$4.4\pm1.0$  
 & $<1.5$\\
Gaussian Process Amplitude (m~s$^{-1}$)& $h$  
 &$5.0\pm1.1$  
 & -\\
 Gaussian Process Correlation Timescale (days)& $\tau$  
 &$14.9^{+12.9}_{-9.4}$  
 & -\\
Gaussian Process Periodicity (days)& $T$  
 &$22.5\pm2.1 $ 
 & -\\
Gaussian Process Weighting & $\Gamma$  
 &$1.5\pm0.3$   
 & -\\
RV Offset (m~s$^{-1}$) & $\gamma$  
 &$7.5\pm3.8$  
 & -\\
RV Jitter (m~s$^{-1}$)& $\sigma_{\rm jit}$  
 &$0.38^{+0.50}_{-0.24}$ 
 & -\\
 RV Drift (m~s$^{-1}$~day$^{-1}$) & $\Dot{\gamma}$ 
 &$-0.0017\pm0.0008$  
 & -\\
\hline
{\bf Derived Parameters}\\
Planetary Radius ($R_\oplus$)  & $R_{\rm p}$  
&$2.268\pm0.069$
 & $1.54\pm0.14$ \\
Planetary Mass ($M_\oplus$)  & $M_{\rm p}$ 
 & $9.2\pm2.1$
 & $<3.6$\\
\enddata
\end{deluxetable*}

\begin{figure*}
\center
\includegraphics[width = 1.5\columnwidth]{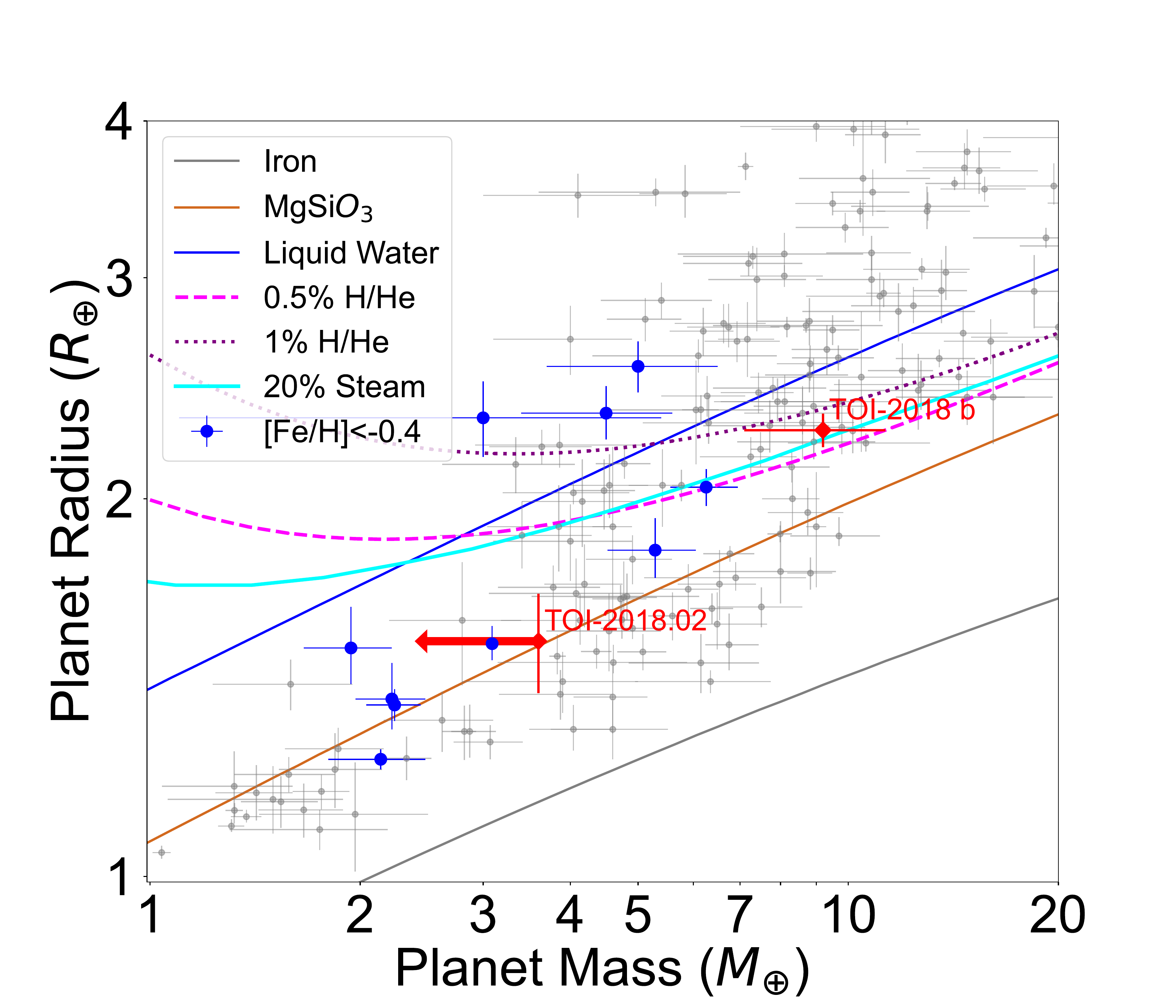}
\caption{The mass and radius measurement of TOI-2018 b and TOI-2018.02 (upper limit only, low-SNR detection ). The solid curves are theoretical mass-radius curves from \citet{Zeng2016}. The dotted and dashed lines are mass-radius curves for planets with an Earth-like core and a H/He envelope of varying mass ratios \citep{Chen_Rogers}. We also showed the updated mass-radius curves for water worlds when the supercritical state of water is taken into account in the equation of state \citep[cyan curves shows a 20\%-by-mass water layer on top of an Earth-like core,][]{Aguichine}. We highlight other planets around low-metallicity host stars (blue symbols; [Fe/H]<-0.4). Even super-Earths \citep[$<$1.5$R_\oplus$,][]{Rogers} planets around low-metallicity stars seem to favor a lower mean density.}
\label{fig:mass_radius}
\end{figure*}

\begin{figure}
\center
\includegraphics[width = 1.\columnwidth]{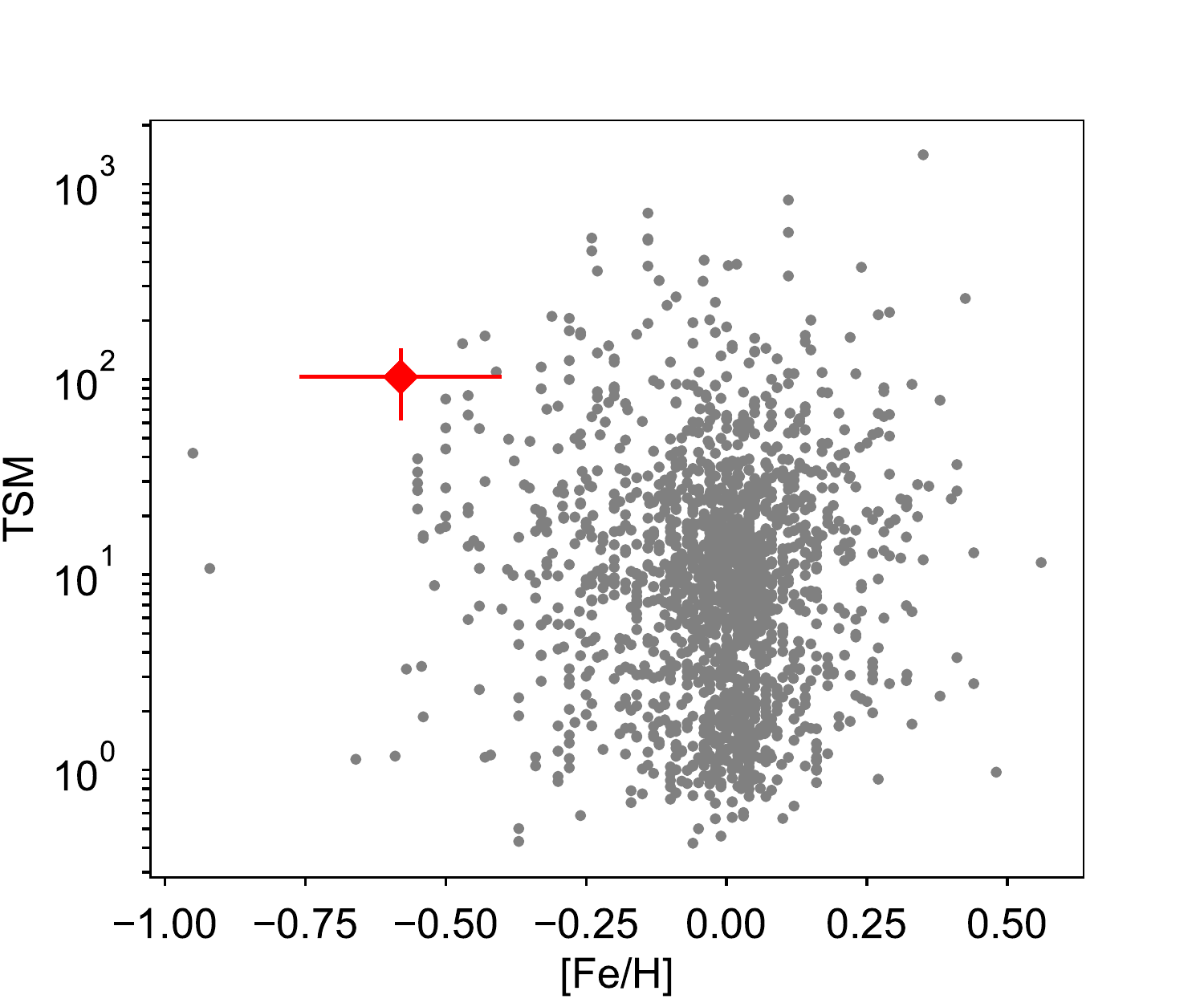}
\caption{The host metallicity [Fe/H] versus the Transmission Spectroscopy Metric \citep[TSM][]{Kempton} of TOI-2018 b (red) and confirmed exoplanets with a mini-Neptune size ($<3R_\oplus$). TOI-2018 b has an estimated TSM of 103.}
\label{fig:tsm}
\end{figure}

\section{Discussion}
In Fig. \ref{fig:mass_radius}, we plot the measured masses and radii of the TOI-2018 planets and other confirmed exoplanets from the NASA Exoplanet Archive\footnote{\url{https://exoplanetarchive.ipac.caltech.edu}}. We also show various theoretical mass-radius relationships from \citet{Zeng2016} including 100\%-Fe, 100\%-MgSiO$_3$, and 100\%-H$_2$O. In addition, we used the model by \citet{Chen_Rogers} to generate the mass-radius relationships of planets with an Earth-like core and a H/He envelope of 0.5\% and 1\% in mass, taking into account the age and insolation of the planet as well. TOI-2018 b lies between 0.5\% and 1\% of H/He. However, TOI-2018 b is also consistent with an ice-rock mixture (H$_2$O-MgSiO$_3$). If we adopt a simple two-layer
model \citep{Zeng2016}, TOI-2018 b is consistent with a 50\%H$_2$O-50\%MgSiO$_3$ composition, with a large uncertainty of about 50$\pm$30\% in the water mass fraction. We also showed the updated mass-radius curves for water worlds when supercritical water is included in the equation of state \citep{Aguichine}. TOI-2018 b is consistent with having a 20\%-by-mass water/steam layer on top of an Earth-like core when the supercritical state is accounted for. With only mass and radius measurements, one can not distinguish between a H/He enveloped TOI-2018 b from a water-world TOI-2018 b. This ambiguity in composition is common to many exoplanets, which is why it has been difficult to resolve the ongoing debate over whether the observed bimodal radius distribution for sub-Neptune planets \citep{Fulton} is due to the atmrospheric erosion of H/He envelopes \citep[e.g.][]{Owen,OwenWu,Lopez2014,Ginzburg} or the core growth model that gives rise to water worlds \citep[e.g.][]{Zeng2019,Luque,Piaulet}. As noted by \citet{Luque}, late-type stars may be well suited to settle this debate. Water worlds are erioxpected to form beyond the snowline in the disk before migrating to the close-in orbits we see them in today. For late-type stars, the snowlines are generally much closer to the host star \citep{Kennedy}. Moreover, Type-I migration proceeds faster for planets with a higher planet-to-star mass ratio \citep[see ][and references therein]{Kley_2012}. The short migration scale and faster Type-I migration rate both favor the migration of water worlds toward the late-type star. Disk migration might have deposited a sample of close-in water worlds (typically 50\%H$_2$O-50\%MgSiO$_3$) around late-type host stars as reported by \citet{Luque}. The composition of TOI-2018 b is consistent with a water-world interpretation. If future follow-up observations could confirm TOI-2018.02 and its near-resonant (1\% wide of 3:2) configuration with TOI-2018 b, it would further support the hypothesis that the planets underwent inward migration. This is because Type-I migration is a primary channel for capturing planets into mean-motion resonances \citep[e.g.][]{Kley_2012,Batygin2015_capture,Macdonald2018}. \citet{Brewer} reported evidence that the fraction of compact, multi-planet systems is enhanced around low-metallicity host stars. If TOI-2018.02 can be confirmed by future observations, TOI-2018 presents another example of such an orbital architecture around a low-metallicity star.

With a mass of $9.2\pm2.1M_\oplus$, TOI-2018 b is close to the threshold for run-away accretion and hence giant planet formation \citep{Rafikov,LeeChiang2015,Lee2019}. Moreover, the low-metallicity (hence low opacity) envelope of the planet should have cooled more easily and facilitated further accretion \citep{LeeChiang2015}. Within the validity of these models, it might seem strange that TOI-2018 b failed to undergo run-away accretion given the large core mass. Could this somehow be related to the suppressed occurrence rate of gas giant planets around low-metallicity host stars \citep{Fischer2005}? One possible explanation is that the disk lifetime is much shorter around lower-metallicity stars, as suggested by \citet{Yasui}. Theoretically, the efficiency of the photoevaporation (and dissipation) of protoplanetary disks is enhanced at lower metallicity \citep[the timescale for disk photoevaporation $t_{\rm phot}\propto Z^{0.52}$][]{Ercolano}. See also more recent hydrodynamic simulations by \citet{Nakatani}. It may be the case that TOI-2018 b did not have enough time to initiate run-away accretion before the disk dissipated. Another relevant work by \citet{Wilson2022} suggested that mini-Neptunes around low-metallicity host stars tend to have higher mean densities. It may indeed be the case that mini-Neptunes around low-metallicity stars typically does not accrete a thick envelope before the disk dissipates.  

One measurement that may distinguish a planet with H/He envelope and a water world is to look for metastable Helium absorption due to the exosphere of the planet in the near infrared \citep[e.g.][]{Oklop2017,Spake}. K-type stars are ideal targets for metastable Helium observation thanks to their balance of extreme UV to far UV flux which respectively excites and destroys the metastable He population \citep{Oklop2019,WangDai69}. A substantial  metastable He population in turn leads to the absorption of the 10830 \AA~~transition. With a J-band magnitude of 7.8 and a moderately active K-type host, TOI-2018 b is a favorable target to look for metastable Helium absorption. The detection of ongoing Helium escape would strongly favor a H/He envelope that has survived photoevaporation.

We quantify the observability of TOI-2018 b in transmission spectroscopy using the James Webb Space Telescope \citep[JWST,][]{Gardner}. We computed the Transmission Spectroscopy Metric (TSM) as suggested by \citet{Kempton}. TOI-2018 b has a TSM of roughly 103. Although there are dozens of known exoplanets that have a higher TSM (Fig. \ref{fig:tsm}), TOI-2018 b does provide a rare opportunity to probe the atmospheric composition of planets formed in a low-metallicity environment. It is one of the top-ranking TSM targets with [Fe/H]<-0.5. Given how bright the host is (J=7.8, K=7.1), special attention of the choice of instruments and observation modes is required to avoid saturation.

Previous results by \citet{Brinkman} and \citet{Demangeon} may suggest that super-Earths formed around low-metallicity late-type stars (L 98-59 M dwarf [Fe/H] = -0.46$\pm0.26$ and TOI-561  K dwarf [Fe/H] = -0.41$\pm0.05$) have lower mean densities than super-Earths around Sun-like stars. A similar trend was also pointed out by \citet{Adibekyan} and \citet{Castro}. The lower mean densities may be the result of an alternative planet formation pathway in the low-metallicity regime. The enhanced $\alpha$-element abundance (Mg, Ca, Si) compared to Fe naturally favors the formation of a larger mantle than an iron/nickel core. If so, one might expect planets around low-metallicity stars (particularly thick disk stars) to have lower mean densities compared to solar-type stars.  The literature contains only a handful of mass and radius measurements for planets around low-metallicity host stars ([Fe/H] < -0.4). More metal-poor systems and more precise characterization of these planets are needed to evaluate their composition as a population.

\software{{\sc AstroImage} \citep{Collins:2017}, {\sc Isoclassify} \citep{Huber},{\sc isochrones} \citep{Morton} {\sc MIST} \citep{MIST}, {\sc SpecMatch-Syn} \citep{Petigura_thesis}, {\sc Batman} \citep{Kreidberg2015}, {\sc emcee} \citep{emcee}, {\sc iSpec} \citep{blanco-cuaresma2014,blanco-cuaresma2019}, {\sc colte} \citep{casagrande2021}}

\facilities{Keck:I (HIRES), {\it TESS},MuSCAT2, WASP, Palomar, Lick, Gemini, Carlo ALto, Caucasian Observatory of Sternberg Astronomical Institute}

\begin{acknowledgments}
\begin{center}
ACKNOWLEDGEMENTS
\end{center}
We thank Heather Knutson, Yayaati Chachan, Kento Masuda, and Yanqin Wu for helpful discussion. 

This material is based on work supported by the TESS General Investigator program under NASA grant 80NSSC20K0059.

D.H. acknowledges support from the Alfred P. Sloan Foundation, the National Aeronautics and Space Administration (80NSSC21K0652), and the Australian Research Council (FT200100871).

J.M.A.M. is supported by the National Science Foundation (NSF) Graduate Research Fellowship Program under Grant No. DGE-1842400. J.M.A.M. acknowledges the LSSTC Data Science Fellowship Program, which is funded by LSSTC, NSF Cybertraining Grant No. 1829740, the Brinson Foundation, and the Moore Foundation; his participation in the program has benefited this work.

This article is based on observations made with the MuSCAT2 instrument, developed by ABC, at Telescopio Carlos S\'{a}nchez operated on the island of Tenerife by the IAC in the Spanish Observatorio del Teide. This work is partly financed by the Spanish Ministry of Economics and Competitiveness through grants PGC2018-098153-B-C31.

This work is partly supported by JSPS KAKENHI Grant Number JP18H05439
and JST CREST Grant Number JPMJCR1761.
This article is based on observations made with the MuSCAT2
instrument, developed by ABC, at Telescopio Carlos Sánchez operated on
the island of Tenerife by the IAC in the Spanish Observatorio del
Teide.

DRC acknowledges partial support from NASA Grant 18-2XRP18\_2-0007. 

AAB acknowledges the support of Ministry of Science and Higher Education of the Russian Federation under the grant 075-15-2020-780 (N13.1902.21.0039).

The data presented herein were obtained at the W. M. Keck Observatory, which is operated as a scientific partnership among the California Institute of Technology, the University of California and the National Aeronautics and Space Administration. The Observatory was made possible by the generous financial support of the W. M. Keck Foundation.

The authors wish to recognize and acknowledge the very significant cultural role and reverence that the summit of Maunakea has always had within the indigenous Hawaiian community.  We are most fortunate to have the opportunity to conduct observations from this mountain.

We acknowledge the use of public TESS data from pipelines at the TESS Science Office and at the TESS Science Processing Operations Center. Resources supporting this work were provided by the NASA High-End Computing (HEC) Program through the NASA Advanced Supercomputing (NAS) Division at Ames Research Center for the production of the SPOC data products.

J. L.-B. acknowledges financial support from the Spanish Ministerio de Ciencia e Innovacion (MCIN\/AEI\/ 10.13039\/501100011033) and the European Union NextGeneration EU/PRTR under the Ramony Cajal program with code RYC2021\-031640\-I. Based on observations collected at the Centro Astronomico Hispano en Andalucia (CAHA) at Calar Alto, operated jointly by Junta de Andalucia and Consejo Superior de Investigaciones Cientificas (IAA-CSIC).

A.C.-G. is funded by the Spanish Ministry of Science through MCIN/AEI/10.13039/501100011033 grant PID2019-107061GB-C61.

This paper made use of data collected by the TESS mission and are publicly available from the Mikulski Archive for Space Tele- scopes (MAST) operated by the Space Telescope Science Institute (STScI). 
 
Funding for the TESS mission is provided by NASA’s Science Mission Directorate.

Some of the observations in this paper made use of the High-Resolution Imaging instrument ‘Alopeke and were obtained under Gemini LLP Proposal Number: GN/S-2021A-LP-105. ‘Alopeke was funded by the NASA Exoplanet Exploration Program and built at the NASA Ames Research Center by Steve B. Howell, Nic Scott, Elliott P. Horch, and Emmett Quigley. Alopeke was mounted on the Gemini North telescope of the international Gemini Observatory, a program of NSF’s OIR Lab, which is managed by the Association of Universities for Research in Astronomy (AURA) under a cooperative agreement with the National Science Foundation. on behalf of the Gemini partnership: the National Science Foundation (United States), National Research Council (Canada), Agencia Nacional de Investigación y Desarrollo (Chile), Ministerio de Ciencia, Tecnología e Innovación (Argentina), Ministério da Ciência, Tecnologia, Inovações e Comunicações (Brazil), and Korea Astronomy and Space Science Institute (Republic of Korea).

\end{acknowledgments}

\bibliography{main}

\begin{deluxetable}{ccccc}
\tabletypesize{\scriptsize}
\label{tab:rv}
\tablecaption{Keck/HIRES Radial Velocities }
\tablehead{
\colhead{Time (BJD)} & \colhead{RV (m/s)} & \colhead{RV Unc. (m/s)} & \colhead{$S_{HK}$}& \colhead{$S_{HK}$ Unc.} }
\startdata
2455730.958602 & 6.07 & 1.45 & 0.862 & 0.001\\
2455734.825559 & 8.22 & 1.28 & 0.924 & 0.001\\
2455738.787684 & 4.87 & 1.24 & 0.942 & 0.001\\
2459038.821696 & 13.18 & 1.17 & 0.980 & 0.001\\
2459040.797956 & 1.46 & 1.28 & 0.943 & 0.001\\
2459041.824545 & -2.66 & 1.16 & 0.976 & 0.001\\
2459057.899283 & 1.45 & 1.48 & 0.923 & 0.001\\
2459379.877596 & 10.84 & 1.29 & 0.955 & 0.001\\
2459381.996691 & -1.29 & 1.35 & 0.939 & 0.001\\
2459383.007785 & -3.82 & 1.48 & 0.908 & 0.001\\
2459383.838423 & -8.58 & 1.22 & 0.930 & 0.001\\
2459384.840558 & -11.87 & 1.27 & 0.966 & 0.001\\
2459385.899575 & -10.13 & 1.30 & 0.909 & 0.001\\
2459386.771843 & -5.64 & 1.17 & 0.940 & 0.001\\
2459388.873567 & -5.48 & 1.33 & 0.920 & 0.001\\
2459389.835862 & -4.30 & 1.42 & 0.905 & 0.001\\
2459395.8831 & 1.87 & 1.23 & 0.895 & 0.001\\
2459397.962232 & -0.29 & 1.49 & 0.873 & 0.001\\
2459399.786738 & -2.54 & 1.24 & 0.902 & 0.001\\
2459404.959869 & 4.76 & 1.14 & 0.922 & 0.001\\
2459405.967385 & -2.93 & 1.26 & 0.949 & 0.001\\
2459406.835315 & -1.98 & 1.15 & 0.935 & 0.001\\
2459408.949042 & 1.34 & 1.18 & 0.907 & 0.001\\
2459409.954172 & 3.36 & 1.14 & 0.873 & 0.001\\
2459411.773495 & -0.61 & 1.41 & 0.876 & 0.001\\
2459412.953324 & -2.68 & 1.32 & 0.874 & 0.001\\
2459414.954138 & 1.18 & 1.44 & 0.888 & 0.001\\
2459422.908053 & 3.78 & 1.51 & 0.889 & 0.001\\
2459435.803746 & -3.35 & 1.23 & 0.854 & 0.001\\
2459443.768482 & -4.01 & 1.16 & 0.857 & 0.001\\
2459444.854338 & -4.55 & 1.14 & 0.875 & 0.001\\
2459446.857033 & 13.66 & 1.54 & 0.928 & 0.001\\
2459448.79132 & 11.51 & 1.18 & 0.916 & 0.001\\
2459449.79849 & 6.07 & 1.23 & 0.925 & 0.001\\
2459451.81223 & -3.59 & 1.11 & 0.960 & 0.001\\
2459452.74112 & -1.93 & 1.11 & 0.945 & 0.001\\
2459455.773054 & -3.80 & 1.26 & 0.886 & 0.001\\
2459456.806188 & -3.19 & 1.27 & 0.890 & 0.001
\enddata
\end{deluxetable}

\end{document}